\documentclass[12pt]{article}

\newif\ifpdf
\ifx\pdfoutput\undefined
\pdffalse 
\else
\pdfoutput=1 
\pdftrue
\fi

\ifpdf
\usepackage[pdftex]{graphicx}
\else
\usepackage{graphicx}
\fi

\usepackage{setspace}
\newcommand{\e}{\mbox{e}}

\title{
{\normalsize \hfill SPIN-2002/11}\\
${}$\\ 
A Hexagon Model for 3D Lorentzian Quantum Cosmology}
\author{B. Dittrich$^1$ and R. Loll$^2$\\
${}$\\
${}^1${\small Max-Planck-Institute for Gravitational Physics,
Am M\"uhlenberg 1, D-14476 Golm}\\
${}^2${\small Institute for Theoretical Physics, Utrecht University,
Leuvenlaan 4, NL-3584 CE Utrecht}}
\begin{document}

\ifpdf
\DeclareGraphicsExtensions{.pdf, .jpg, .tif}
\else
\DeclareGraphicsExtensions{.eps, .jpg}
\fi

\maketitle
\abstract{
We formulate a dynamically triangulated model
of three-dimensional Lorentzian quantum gravity whose
spatial sections are flat two-tori. It is shown that 
the combinatorics involved in evaluating the one-step 
propagator (the transfer matrix) is that of a set of
vicious walkers on a two-dimensional lattice
with periodic boundary conditions and that the entropy
of the model scales exponentially with the volume. 
We also give explicit expressions for the
Teichm\"uller parameters of the spatial slices in
terms of the discrete parameters of the 3d 
triangulations, and reexpress the discretized action
in terms of them. The relative simplicity and
explicitness of this
model make it ideally suited for an analytic study
of the conformal-factor cancellation observed 
previously in Lorentzian dynamical triangulations
and of its relation to alternative, reduced phase
space quantizations of 3d gravity.}

\section{Motivation}\label{mot}
The approach of Lorentzian Dynamical Triangulations\footnote{See
\cite{amb,loll} for recent reviews.}
(LDT) leads to a well-defined regularized path integral for 
3d quantum gravity as was shown in \cite{ajl1,d3d4}. The phase
structure of this statistical model of causal random
geometries has been investigated by Monte Carlo methods in 
the genus-zero case, where the two-dimensional spatial slices
are spheres \cite{ajl2,ajl3}. Its maybe most striking feature
is the emergence in the continuum limit of 
a well-defined ground state behaving macroscopically
like a three-dimensional universe \cite{ajl2,adjl,ajl4}. 
This is in contrast with perturbative continuum arguments 
which suggest that in $d\geq 3$ Euclideanized gravitational
path integrals are generically ill-defined because of a
divergence due to the conformal mode. Since a Wick rotation
from Lorentzian to Euclidean space-time geometries is 
part of the evaluation of the regularized state sums in LDT, 
one might expect to encounter a similar problem here, but this
is not what happens. Instead, all indications point to a 
non-perturbative cancellation between the conformal
term in the action (which still has the same structure as in the
continuum) and entropy contributions to the state sum
(that is, ``the measure'').
It should also be emphasized that this cancellation is not
achieved by any ad hoc manipulations of the path integral,
for example, by isolating the conformal mode and Wick-rotating
it in a non-standard way (in fact, it is quite impossible to
isolate this mode in the non-perturbative setting of LDT). --
Further discussions of the conformal-mode problem and
its possible non-perturbative resolution can be found in 
\cite{conf,adjl}.

It is obviously of great interest to understand in a more
explicit and analytic fashion how this cancellation occurs and
how it gives rise to an effective Hamiltonian whose ground
state is the one seen in the numerical simulations of
3d Lorentzian dynamical triangulations.

Some progress in this direction has been made recently 
by mapping the
three-dimensional LDT model to a two-dimensional 
Hermitian ABAB-matrix model \cite{ajlv,kz}. This latter model has a
second-order phase transition which is absent from the
LDT model. This comes about because the matrix model
naturally contains generalized geometric configurations 
which are not allowed in the original quantum gravity
model, and which can be interpreted as wormhole geometries.
The second-order transition is related to the
abundance of such wormholes and the original LDT model
corresponds to the weak-gravity phase of the matrix
model below the critical value of Newton's constant.\footnote{
The reason why a {\it two}-dimensional matrix model 
appears in the description of {\it three}-dimensional
quantum gravity is the fact that the 3d space-time
geometries of the latter can be uniquely characterized
by a sequence of 2d graphs representing the intersection
patterns of the 3d triangulations at constant half-integer
times $t+1/2$.}

Although the mapping to the matrix model has yielded some 
analytic information about the phase structure of 3d quantum gravity, 
the explicit transfer matrix has not yet been constructed.
This is a desirable goal, because it would lead to a
quantum Hamiltonian that among other things could be 
compared with
already existing canonical quantizations of 3d gravity.
Also, having a more detailed control over the combinatorics 
of the triangulated model would be extremely interesting in
order to understand the precise cancellation mechanism 
between the conformal terms in the action and the entropic
measure contributions. 

In the absence of a solution of the full three-dimensional
model, one strategy is to formulate simplified
models rich enough to capture the dynamics of
3d gravity but whose combinatorial properties at the same 
time are sufficiently simple to allow for an explicit
solution. There are two types of restrictions one can
naturally impose on discretized Lorentzian space-times.
The first are restrictions on the allowed spatial
geometries (these are two-dimensional simplicial manifolds
built from equilateral Euclidean triangles) at integer 
proper times $t$,
and the second are restrictions on the allowed 
three-geometries that interpolate between adjacent
spatial slices at times $t$ and $t+1$. Note that the
first type of restriction has a direct influence on
the Hilbert space of the system, since the spatial
geometries may be thought of as a basis in the
position-representation (where ``position'' here stands for a
spatial geometry). 

Such restrictions may or may not change the critical
properties of the corresponding ensemble of random
geometries. In a larger context, a task that still needs to
be accomplished is the classification 
of all possible three-dimensional LDT models according
to their critical properties and dynamics as a
function of the set of Lorentzian discretized
space-times allowed in the state sum. (As already
mentioned earlier, one may not only consider imposing
restrictions on the class of geometries, but may also allow
for generalizations.) In general, one would expect this
large number of possible discrete models to fall
into only a small number of universality classes. 
This turned out to be the case in two dimensions,
where there are essentially two universality classes,
depending on whether one allows space-time to grow
so-called ``baby-universes'' or not \cite{al} (however,
see \cite{charlotte} for some ``exotic'' variations). 
Of course, three-dimensional models of random geometries
have been studied far less, and it is not a priori
clear what structures one should expect to find.

A number of ``quantum-cosmological'' LDT models of
2+1 gravity were considered in \cite{dehne}, see also 
the discussion in \cite{adjl}.
There, the number of three-geometries contributing to the path
integral was restricted by imposing symmetry constraints,
reflecting homogeneity and isotropy properties of space.
The space-time topology was fixed to $I\times T^2$, that is,
with toroidal spatial slices. Imposing in addition
(discrete) spatial translation invariance fixes the
tori at integer-$t$ to be locally flat and Euclidean.
The spatial geometries are then completely characterized 
by three numbers, namely, the torus volume and its
(two real) Teichm\"uller parameters. The reason for why
one may still hope to capture the essential dynamical
features of 3d quantum gravity this way -- despite the 
drastic reduction in the degrees of freedom -- is 
that canonical continuum considerations suggest that 3d
quantum gravity has only a finite number of true physical 
degrees of freedom (which are precisely the
Teichm\"uller parameters). Note also that the torus case
is the simplest choice with non-trivial Teichm\"uller
parameters and also the one which has been most
studied in the literature \cite{2+1}.

In the simplest and most restrictive model of such 
a torus universe one demands that also the
spatial intersections at constant half-integer $t$ 
should be lattice-translationally invariant \cite{dehne}
(see \cite{jormaetal} for related cosmological continuum
models).
This scenario is most easily implemented by choosing
as fundamental building blocks tetrahedra and pyramids; see
\cite{kappel} for a generalization to 3+1 dimensions.
Although it is straightforward to work out the combinatorics
of all possible interpolating three-geometries between
two spatial slices, it turns out that the model does not
possess an interesting continuum limit. This has to do with
the fact that because of the strong symmetry restrictions 
there are too few ``microstates'', that is, too few geometries
contributing to the state sum at any given value of the
action. More specifically, unlike in 3d LDT without such
restrictions, the number of distinct triangulations
of a given space-time slice $\Delta t=1$ (corresponding to a
single time step) grows only exponentially $\sim {\rm e}^{const.\, L}$ 
with the linear size $L$ of the spatial torus. Since this entropy
term has to compete with the volume-suppressing 
exponentiated cosmological term from the Euclideanized
action which is of the form ${\rm e}^{-\lambda \Delta t\, L^2}
={\rm e}^{-\lambda\, L^2}$, the state sum will always be
dominated by geometries with effectively one-dimensional
spatial slices as the number of tetrahedral building blocks
goes to infinity. Any potential continuum limit is therefore 
unlikely to have anything to do with the original LDT model, 
and we must conclude 
that this cosmological model is simply not rich enough to
study the conformal-factor cancellation and the effective
quantum dynamics of three-dimensional quantum gravity.
(For a general discussion of renormalization
and continuum limits in dynamical triangulations approaches
to gravity see reference \cite{amb}.)

In the present piece of work, we will investigate an alternative, less 
restrictive cosmological LDT model first introduced in \cite{dittrich}. 
Its spatial two-geometries are still 
given by flat tori, but the symmetry restrictions on the interpolating 
space-time geometries are relaxed. We describe this so-called
hexagon model in Sec.\ \ref{hex}. As usual, a given discretized
space-time contributing to the propagator consists of a sequence 
of layers $[t,t+1]$. For the hexagon model, the geometry of each 
such ``sandwich" can be characterized as a tesselation
of a regular 2d triangular lattice by coloured rhombi. The two-coloured
graph dual to this tesselation is a superposition of two regular
hexagonal graphs describing the flat two-tori which form the
space-like boundary of the sandwich. In Sec.\ \ref{act} we compute
the Lorentzian action for a sandwich geometry, together with
its Euclidean counterpart. 

Our next task is the counting of all possible interpolating sandwich
geometries for given torus boundaries. We show in Sec.\ \ref{comb} 
that the associated combinatorics is that of a set of ``vicious
walkers'' on a 2d lattice with periodic boundary conditions. In
the following section, we demonstrate that the contribution
$\Delta S$ to the action of a single sandwich is already essentially
determined by the geometry of the flat tori which form its
boundary. Still, there is a large number of microstates for given
boundary data, and we prove that their number indeed grows
to leading order exponentially with the torus {\it volume}, and not 
just linearly. In Sec.\ \ref{teich} we calculate explicitly the variables
describing the flat spatial tori (for each torus, two real Teichm\"uller 
parameters and the two-volume) in terms of the 
data labelling a triangulated space-time sandwich. 
Since the latter are a set of discretized variables, it is of interest
to see how they sample the usual continuous Teichm\"uller space of
all flat tori. This is illustrated in Sec.\ \ref{sample} by explicitly 
calculating the Teichm\"uller parameters for a set of
geometries whose volume is smaller than a certain cutoff.
We also include a sample plot of the
associated moduli space, obtained by factoring out the
large diffeomorphisms. Our conclusions are then presented in
Sec.\ \ref{concl}. -- Appendix 1 contains details of the coordinate
transformation between the discrete geometric parameters
and the torus data, and Appendix 2 an asymptotic evaluation
of the vicious-walker combinatorics relevant to the
entropy estimate.

\section{The hexagon model}\label{hex}

The fundamental three-dimensional building blocks used
in this model are (in the language of \cite{d3d4}) 
3-1 tetrahedra, glued together pairwise, 1-3 tetrahedra, also
glued together pairwise, and
single 2-2 tetrahedra. The numbers $i$-$j$ indicate that
a tetrahedron shares $i$ vertices with the spatial
geometry at time $t$ and $j$ vertices with that at time $t+1$.
As usual, the space-like edges of a tetrahedron all have
squared length $1$ (or $a^2$ in units of the lattice
spacing $a$) and the time-like edges squared length
$-\alpha$ (or $-\alpha a^2$), where $\alpha >0$ is real.
The pairing of the 3-1 or 1-3 tetrahedra is obtained
by gluing them along a (time-like) triangular face.

\begin{figure}[t]
\centerline{\scalebox{0.6}{\rotatebox{0}
{\includegraphics{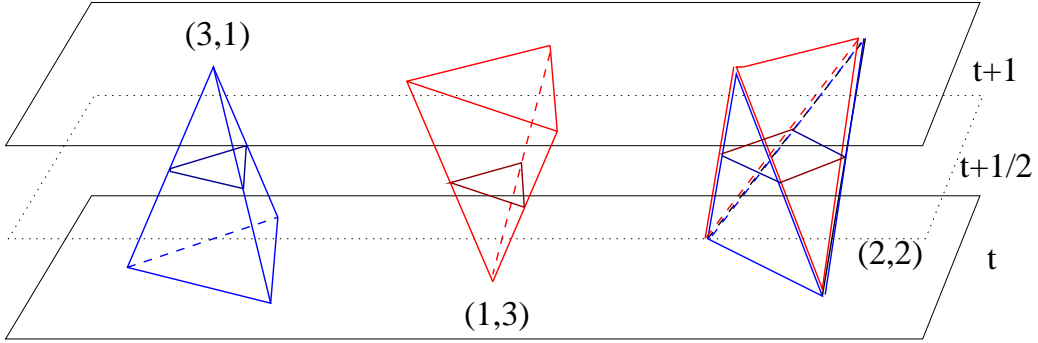}}}}
\caption[blocks]{The three types of tetrahedral
building blocks and their corresponding two-dimensional
intersection patterns at $t=1/2$. 
By definition of the model, both the 3-1 and the 1-3 tetrahedra
always occur in pairs.}
\label{blocks}
\end{figure}

The three types of building blocks are illustrated in
Fig.\ref{blocks}. Any three-dimensional ``sandwich
geometry'' of height $\Delta t=1$ we construct from 
these building blocks can be uniquely described by
the intersection pattern that results when the
tetrahedra are sliced in half at time $t+1/2$ and the
time-like triangles that are cut in the process are
represented by one-dimensional links. We colour-code the
links to distinguish whether they come from triangles
with tip at $t$ (red links) or tip at $t+1$ (blue links).
Our three building blocks can thus be represented by
blue and red double-triangles and by squares with
alternating blue-red sides. Topologically, the intersection
graph is again a torus.

\begin{figure}[h]
\centerline{\scalebox{0.45}{\rotatebox{0}
{\includegraphics{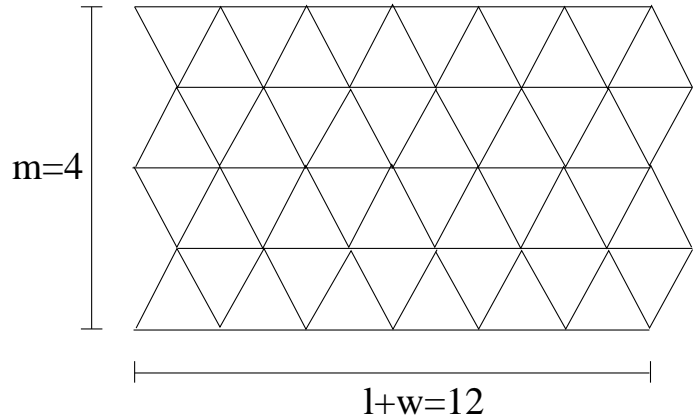}}}}
\caption[background]{A regular triangular lattice serving
as a ``background geometry'' at half-integer time $t+1/2$.}
\label{background}
\end{figure}

As already mentioned, we want to consider only amplitudes
between {\it flat} two-tori. For the blue-red intersection
picture this means that when the red links are
simultaneously shrunk to zero length, what must remain
is a regular tiling of a torus by (blue) triangles,
where exactly six triangles meet at each vertex (and
similarly at time $t+1$ when we shrink away the blue
links). A systematic way of generating intersection
patterns with this property is as follows. Take a
rectangular strip of a regular triangular lattice
of discrete width $l+w$ and height $m$, where the units
are chosen in such a way that all vertices have
integer coordinates\footnote{The reason for splitting up the
width into two integers will become clear in Sec.\ \ref{comb}.} 
(Fig.\ref{background}). Let us for the moment
take $l+w$ and $m$ to be even, $l=2l'$, $w=2w'$, 
$m=2m'$, $l',w',m'\in Z_+$, since 
this will make it possible to identify
the opposite sides of this strip without any twists to
create a compact two-torus. (In a more general model,
one may also allow for twists in either of these
directions.)

\begin{figure}[h]
\centerline{\scalebox{0.6}{\rotatebox{0}
{\includegraphics{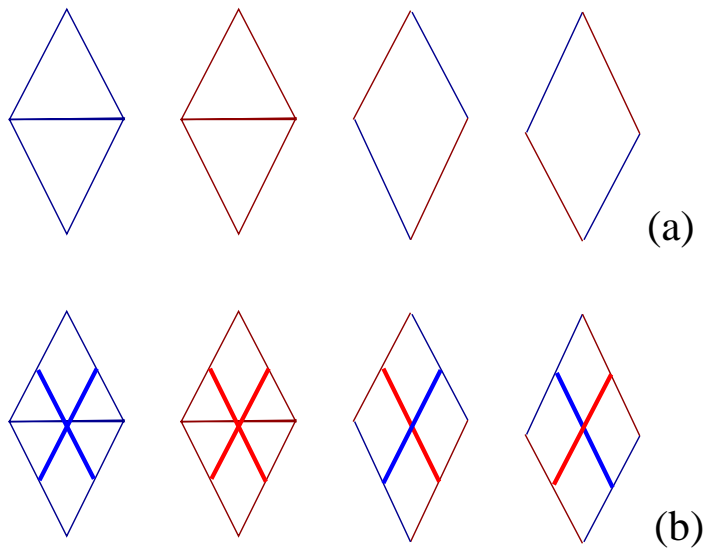}}}}
\caption[rhombi]{The four types of rhombic tiles (a),
and their dual representation (b).}
\label{rhombi}
\end{figure}

This regular lattice is to serve as a ``background geometry''
which we are going to tile with rhombic 2d building blocks
so that no space is left blank. It is immediately clear
that the intersections of the double-tetrahedra
are blue and red rhombi. The squares that result from
cutting the 2-2 tetrahedra will be ``distorted'' in this
representation so that they can fit onto the triangular
lattice. Since this can be done in two ways, we have
a total of four rhombic tiles at our disposal (Fig.\ref{rhombi}a). 
It will be convenient for our purposes to
adopt a dual notation where each rhombus is represented
by a pair of crossing links. Each link connects two
opposite edges of the rhombus and has the same colour,
see Fig.\ref{rhombi}b. The rhombi can only be put onto the 
lattice if the colours of their edges (or their dual links)
match pairwise at intersections. 

The beautiful feature of this representation is the
fact that any tiling of the strip (which of course
must be compatible with its periodic identifications)
automatically leads to in- and out-geometries which
are flat, connected tori. The easiest way of seeing this
is by following dual links (or pieces of dual links) of
a given colour around a closed loop, where the loop
must be such that no more lines of the same colour
branch off into the loop's interior. Next, consider how this loop
is represented in terms of triangles of the same
colour which make up one of the adjacent spatial geometries.
The pieces of straight coloured lines coming from
the last two building blocks depicted in Fig.\ref{rhombi}b
do not correspond to any triangles at all. By contrast,
the first two rhombic tiles correspond to a couple of
adjacent triangles each in the relevant spatial geometry
at integer $t$. Since the rhombic tiles can be put onto
the triangular ``background lattice'' at half-integer $t$
only with three possible orientations (see Fig.\ref{abc} 
below),
it is clear that any of the loops introduced above
corresponds to a sequence of exactly six triangles.

\begin{figure}[t]
\centerline{\scalebox{0.6}{\rotatebox{0}
{\includegraphics{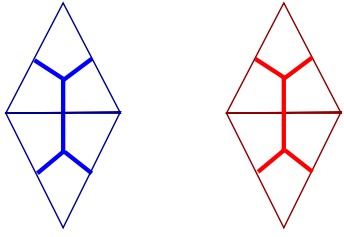}}}}
\caption[twotriangle]{These tiles appear as blue and
red double-triangles in spatial slices of integer $t$.
Their trivalent dual graphs form part of regular
hexagonal graphs representing flat two-geometries.}
\label{twotriangle}
\end{figure}

In fact, these two types of rhombi can more properly
be represented by dual trivalent graphs which are
dual to the individual triangles, as shown in 
Fig.\ref{twotriangle}.
It is then easy to see that each loop corresponds to
a hexagon graph, with six ``corners'' of 120 degrees each,
a property which has given the model its name.
Translating this into a statement about the 
two-geometry, it means that there are six triangles meeting
at each vertex, hence the geometry is everywhere
flat. Thinking a little further along these lines, one can
also convince oneself that it is not possible to obtain 
a flat triangulation of one colour that consists of two or more
disconnected pieces.

\section{Gravitational action and transfer matrix}\label{act}

The first step in constructing a path integral for
our model is to determine which sandwich geometries
$\Delta t=1$ 
can occur and to compute their contribution to the
action. Following \cite{ajl2}, this action can be written
as a function of the total numbers $N_{31}$, $N_{13}$
and $N_{22}$ of tetrahedral building blocks occurring in
the slice, namely,
\begin{eqnarray}
S(\Delta t\! =\! 1)&=& \nonumber\\
(N_{31}+N_{13})&&\hspace{-1cm}
\Bigl( \pi k \sqrt{\alpha} 
-3 k\ {\rm arcsinh}\ \frac{1}{\sqrt{3}\sqrt{4\alpha +1}}
-3 k \sqrt{\alpha} \arccos \frac{2\alpha+1}{4\alpha +1} -
\frac{\lambda}{12} \sqrt{3\alpha +1}\Bigr)\nonumber \\
\hspace{-.4cm}+N_{22}&&\hspace{-1cm}
\Bigl( 2\pi k \sqrt{\alpha} 
+2 k\ {\rm arcsinh}\ \frac{2\sqrt{2}\sqrt{2\alpha +
1}}{4\alpha +1}
-4 k \sqrt{\alpha} \arccos \frac{-1}{4\alpha+1} -
\frac{\lambda}{12} \sqrt{ 4\alpha +2} \Bigr),\nonumber \\
\label{3dloract}
\end{eqnarray}
where $\lambda$ and $k$ denote the bare cosmological and
inverse Newton's constants.
The positive parameter $\alpha$ appearing in (\ref{3dloract})
describes the ratio between the squared lengths of the
time-like and the space-like edges of the triangulation,
$l_{\rm time}^2=-\alpha l_{\rm space}^2$. 
Since we will evaluate the state sums in the Euclidean
sector of the theory, we need to Wick-rotate all of
our Lorentzian discretized manifolds. As explained in
detail in \cite{d3d4}, this is achieved by continuing
$\alpha$ through the complex lower half-plane to negative 
real values.
Let us for simplicity choose the standard value
$\alpha =-1$, so that all edges have the same length. 
This gives rise to the Euclidean action
\begin{eqnarray}
S^{\rm eu}(\Delta t=1)&=&\!\! (N_{31}+N_{13})\Bigl( 
(-\frac{5}{2}\pi + 6\arccos \frac{1}{3}) k +
\frac{1}{6\sqrt{2}}\lambda \Bigr)\nonumber\\
&& + N_{22} \Bigl( 
(-2\pi + 6\arccos \frac{1}{3}) k +
\frac{1}{6\sqrt{2}}\lambda \Bigr)\nonumber\\
&\equiv& \!\! 
(N_{31}+N_{13})  (-0.468 k +0.118 \lambda)
+N_{22} (1.103 k +0.118 \lambda)\label{theaction}
\end{eqnarray}
Obviously in our model the numbers of building blocks of
type (3,1) and (1,3) are always even. Note also that the
action (\ref{theaction}) contains boundary terms (the
discrete analogues of the usual spatial integrals over the 
extrinsic curvature) in order to make it additive under
gluing of subsequent layers of $\Delta t=1$. The 
action without boundary contributions has the form
\begin{eqnarray}
S^{\rm eu}_{\rm bulk}(\Delta t=1)&=&\!\! (N_{31}+N_{13})\Bigl( 
(-\pi + 3\arccos \frac{1}{3}) k +
\frac{1}{6\sqrt{2}}\lambda \Bigr)\nonumber\\
&& + N_{22} \Bigl( 
(-2\pi + 4\arccos \frac{1}{3}) k +
\frac{1}{6\sqrt{2}}\lambda \Bigr)\nonumber\\
&\equiv& \!\! 
(N_{31}+N_{13})  (0.551 k +0.118 \lambda)
+N_{22} (-1.359 k +0.118 \lambda).\label{bulkaction}
\end{eqnarray}

The partition function or propagator for a
single time-step after the Wick rotation is given by
\begin{equation}
G(g_1,g_2;\Delta t\! =\! 1)=\sum_{T:g_1\rightarrow
g_2} \frac{1}{C(T)}\e^{-S^{\rm eu}(T)},
\label{onestep}
\end{equation}
where the sum is over all possible sandwich
geometries $T$ interpolating between the two
spatial boundary geometries $g_1$ and $g_2$,
and $C(T)$ is the order of the symmetry group
of the triangulation $T$. As usual \cite{d3d4},
expression (\ref{onestep}) defines the transfer
matrix $\hat T$ of the system with respect to the natural
scalar product $\langle g_1|g_2\rangle =\frac{1}{C(g_1)}
\delta_{g_1,g_2}$, via its matrix elements
\begin{equation}
\langle g_2|\hat T|g_1\rangle :=
G(g_1,g_2;\Delta t\! =\! 1).
\label{trans}
\end{equation}

\section{The combinatorics of the intersections}\label{comb}

In trying to characterize all possible intersection patterns
that can occur (that is, all possible 3d sandwich geometries),
it is convenient to break up the combinatorial problem
into two steps. The first one is how to tile the triangular
lattice with (identical) rhombi, and the second how to
introduce a colouring on the resulting tilings (in the form
of drawing chains of dual coloured links onto the rhombi).

\begin{figure}[h]
\centerline{\scalebox{0.45}{\rotatebox{0}
{\includegraphics{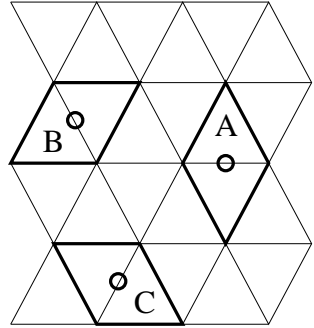}}}}
\caption[abc]{There are three orientations in which
a rhombus can be put onto the triangular background
lattice.}
\label{abc}
\end{figure}

\begin{figure}[h]
\centerline{\scalebox{0.45}{\rotatebox{0}
{\includegraphics{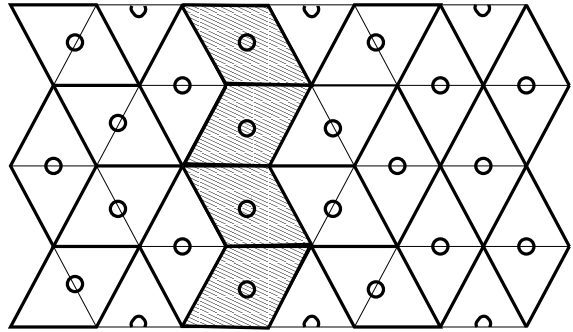}}}}
\caption[tiling]{A tiling with A-, B- and C-rhombi (as usual, 
opposite sides of the strip are to be identified). The shaded
region is a B-C-path of winding number 1 in the vertical
direction.}
\label{tiling}
\end{figure}

A rhombus can be put onto the lattice with three different
orientations, which we call A, B and C. This is illustrated in
Fig.\ref{abc}, where the centres of the rhombi are indicated by
small circles. Fig.\ref{tiling} shows a complete periodic tiling of
a strip with $(l+w,m)=(12,4)$. (Opposite sides of
this strip are to be identified.)
What should be noted here is that
the number of A-blocks as well as the combined number of
B- and C-blocks per horizontal row is conserved as one
advances in steps in the $m$-direction. Starting at some
B- or C-block in row 1, one can therefore follow a ``path'' in vertical
direction made up of some sequence of B- and C-rhombi
until one reaches the upper end of the strip (shaded region in 
Fig.\ref{tiling}). 

\begin{figure}[h]
\centerline{\scalebox{0.4}{\rotatebox{0}
{\includegraphics{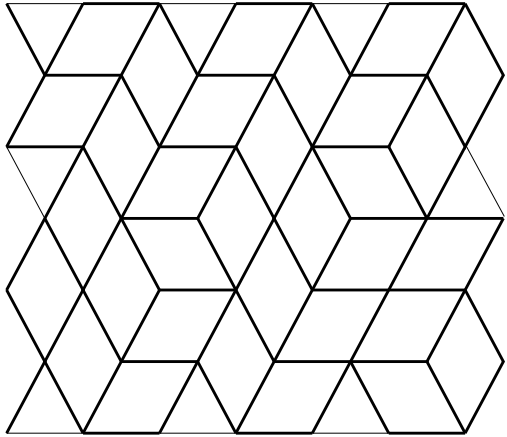}}}}
\caption[forbidden]{An example of a forbidden tiling.}
\label{forbidden}
\end{figure}

At this stage we will for simplicity impose a further
restriction on the allowed patterns of rhombi, namely, that
all B-C-paths should have winding number zero in the
$l$-direction and winding number one in the $m$-direction. 
(This does not seem to impose serious restrictions on
the in- and outgoing two-geometries, c.f Secs.\ \ref{teich},
 \ref{sample}, but is a condition that could be relaxed,
should this turn out to be convenient.) That is, 
following such a path between the lower and upper end of
a strip, it should contain an equal number of B- and
C-rhombi, so that it closes on itself upon identification
of the lower and upper ends of the strip. A configuration
which violates this restriction but is nevertheless
periodic is shown in Fig.\ref{forbidden}. 

\begin{figure}[h]
\centerline{\scalebox{0.5}{\rotatebox{0}
{\includegraphics{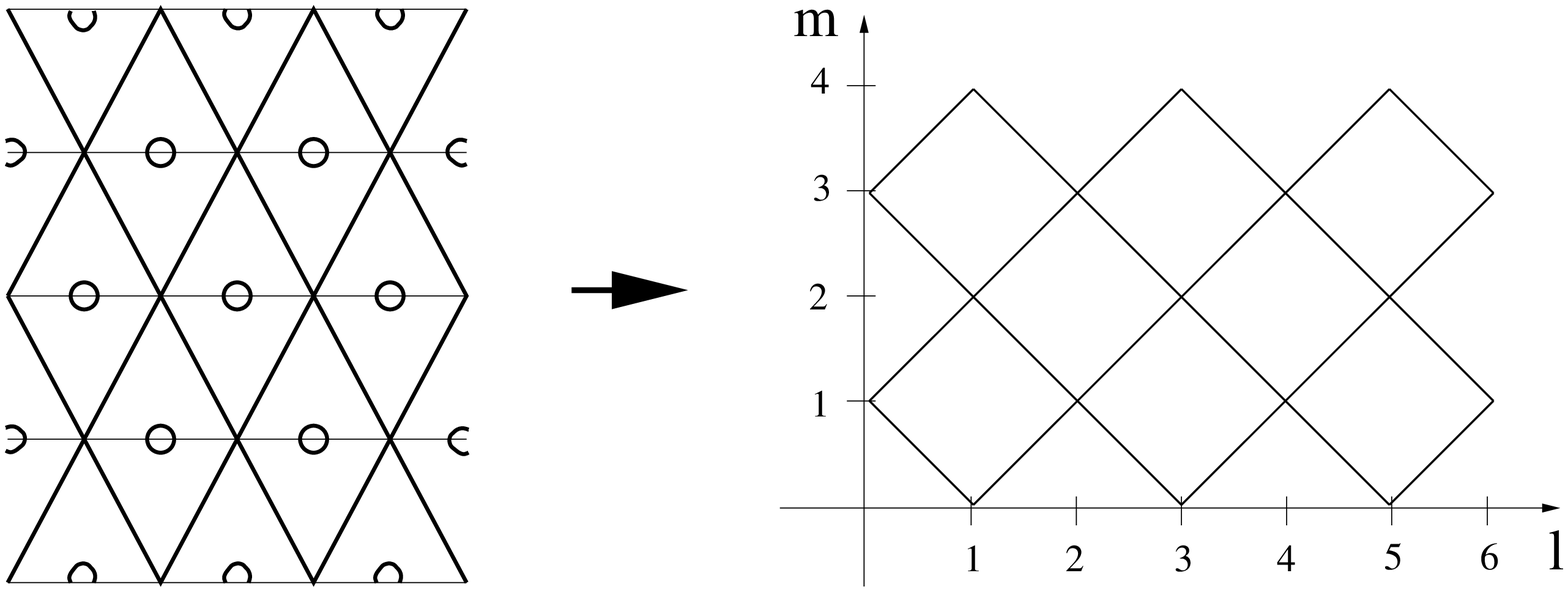}}}}
\caption[astrip]{The regular tiling with A-rhombi obtained
by deleting the B- and C-rhombi from Fig.\ref{tiling} (left),
and the corresponding dual tilted square lattice (right).}
\label{astrip}
\end{figure}

By shrinking all B-C-paths to zero width, one obtains
a regular tiling of only A-rhombi (Fig.\ref{astrip}, left). 
This reduced lattice
can be thought of as a sublattice of the original strip
in the sense that the chains of its dual A-links close onto 
themselves. If the number of B-C-paths was $w/2$, the 
reduced lattice will
have width $l$ and height $m$.
Note also that these chains take the very simple form of 
straight diagonal left- or right-moving lines. 
A configuration of dual A-lines is most easily represented 
as a tilted square lattice (Fig.\ref{astrip}, right). 
The intersection points of the dual left- and
right-moving lines will
then have $l$-coordinate 0,2,4,6,...in the odd rows and
1,3,5,... in the even rows. 

As will become apparent in due course, working out the
possible colour assignments for such dual A-lattices
is an important part of the combinatorics of the model.
They are quite easy to enumerate. By assumption, any
colouring of the dual chains has to respect periodicity
in both $l$- and $m$-directions. The easiest way to
obtain a consistent colouring is therefore as follows.
Start at some (dual) vertex $v$ at $m=0$ and 
colour the, say, left-moving A-line passing through $v$ 
while following it
around the lattice (keeping in mind the periodic
identifications of the strip), until getting back to
the original vertex. Using such a procedure, it is
straightforward to see that the number of  
possible colourings for the entire configuration of
A-lines depends on the integer $d$, the greatest common 
divisor of $l/2$ and $m/2$. 
(For example, the lattice depicted in Fig.\ \ref{astrip}
has $l=6$, $m=4$ and therefore $d=1$.)
Picking an arbitrary vertex $v$ at $m=0$ as origin, 
the colours of the left-
and right-moving A-lines emanating from the first $d$
vertices to its right (including $v$) may be chosen 
arbitrarily, with the remainder determined by periodicity.

To count intersections of a certain type
between the right- and left-moving A-lines (important for
determining the action) it suffices to
look at a fundamental diamond-shaped 
region of $d\times d$ vertices --
this region will then be repeated $lm/2d^2$ times throughout
the lattice. 

\begin{figure}[h]
\centerline{\scalebox{0.5}{\rotatebox{0}
{\includegraphics{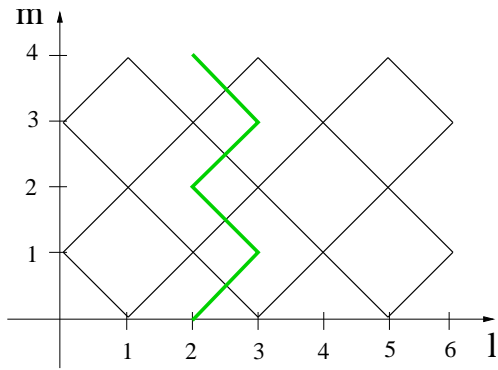}}}}
\caption[bcstrand]{The B-C-strands may be reintroduced
by drawing paths onto the regular dual A-strip. In the
figure, we have drawn the path corresponding to the
shaded area in Fig.\ref{tiling}.}
\label{bcstrand}
\end{figure}

We may now reintroduce the B-C-strands into
this picture by drawing chains of links whose vertices
in every row lie exactly in between the dual A-vertices.
Starting from the initial row $m=0$, we can again follow
paths of dual B- or C-links by making at every vertex a
choice of moving diagonally up to the left or to the right
(Fig.\ref{bcstrand}). More than one dual
B-C-chain can pass through any one vertex, and neighbouring
chains are allowed to share one or more links, but not
to cross, so that their relative position along
the $l$-direction is preserved as we advance in $m$.
(Obviously, to obtain configurations of the type
depicted in Fig.\ref{tiling}, each of these paths must
be ``blown up'' in the horizontal direction to 
width 2.)

The combinatorics of the B-C-chains 
can be mapped onto a model of so-called vicious walkers
(which are {\it not} allowed to touch)
with fixed initial and final points, by inserting columns of 
width 2 between every pair of adjacent B-C-chains\footnote{
{\it Vicious walkers} are imaginary creatures 
that will do {\it vicious} things to each other
when meeting at a point and therefore avoid such
encounters.}. Various versions of vicious-walker models,
differing in their 
boundary conditions for the paths and the underlying lattices,
have been investigated in the literature. 
The most common choice is that of free boundary conditions
in the $l$-direction, i.e. a lattice of effectively infinite width.
The initial points of the walkers at $m=0$ are usually located
at a minimal mutual distance $\Delta l=2$ near the origin,
for example, at $l$-coordinate 0, 2, 4, ... , $w-2$, and the
final points at $m_{\rm max}$ are either chosen freely or 
again grouped together at some point $(l_0,m_{\rm max})$
with $l$-coordinates $l_0$, $l_0+2$, $l_0+4$, ... , $l_0+w-2$
(see, for example, \cite{fisher,essgutt,guttvoe} and references 
therein). An exception is the treatment by Forrester \cite{forr},
who uses periodic boundary conditions in the $l$-direction
and walkers with equally (but not necessarily minimally) 
spaced initial and final positions.

\begin{figure}[h]
\centerline{\scalebox{0.5}{\rotatebox{0}
{\includegraphics{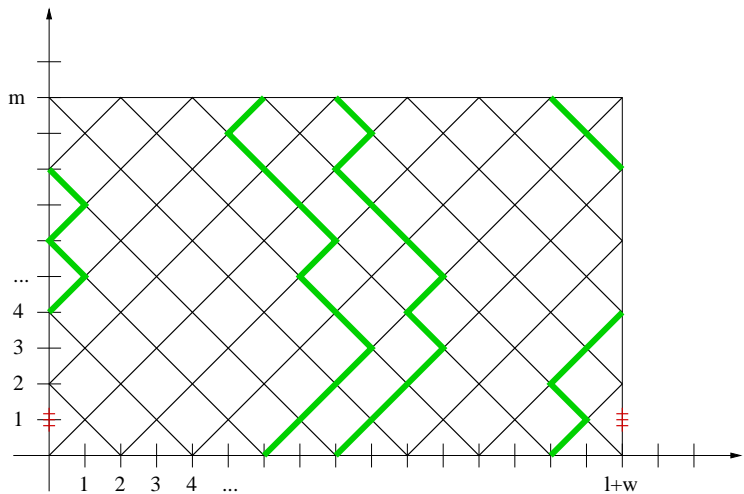}}}}
\caption[walkers]{A configuration of three vicious
walkers, corresponding to $(l,m,w)=(10,10,6)$.}
\label{walkers}
\end{figure}

The case relevant for our 3d gravity model is that of periodic 
boundary conditions in both the $l$- and $m$-direction, and
the combinatorial problem can be phrased as follows:
{\it Given three even integers $l$, $m$ and $w$, 
how many ways are there of drawing $w/2$ indistinguishable
vicious-walker paths with winding numbers $(0,1)$ 
onto a tilted square lattice of 
width $l+w$ and height $m$?} The tilted square lattice
is dual to similar lattices depicted in Figs.\ref{astrip},
\ref{bcstrand},
i.e. it has vertices on the horizontal axis at even parameter values
$0,2,4,\dots,l+w-2$ and vertices on the vertical
axis at values $0,2,4,\dots,m$. Fig.\ref{walkers} shows a typical
configuration of vicious walkers for $(l,m,w)=(10,10,6)$. 
The vertical boundaries of this lattice are identified periodically
as indicated in the figure, so that $l+w\equiv 0$. Each 
vicious-walker path starts on the lower horizontal axis at
some point $(l,0)$ and ends after $m$ steps on the upper
horizontal boundary at the point $(l,m)$, i.e. at the
point with the same horizontal coordinate.

This situation can be viewed as a special case of that of
a single random walker in an alcove of the affine Weyl group
of type $\tilde A_{w/2-1}$ \cite{grab}. The reflections of the
path at the walls of the Weyl chamber correspond in this
case to the collision points of $w/2$ random walkers
that move on a one-dimensional circle. The ensemble of
walkers takes simultaneously steps of unit length along
the circle, either to the right or the left. Mapping
these onto diagonal upward steps on a tilted square lattice
and requiring identical initial and final points for each
walker on the circle leads exactly to a situation as
depicted in Fig.\ \ref{walkers}. Following \cite{gz,grab},
the number of non-intersecting path configurations for
$w/2$ walkers with initial and endpoints
$\vec\lambda=(\lambda_1,\lambda_2,\dots,\lambda_{w/2})$,
$\lambda_i\in\{0,2,4,\dots,l+w-2\}$, ordered along the
circle so that $\lambda_1<\lambda_2<\dots <\lambda_{w/2}$,
is given by
\begin{equation}
b(\vec\lambda,m,l,w)=\sum_{\vec t,\; \sum t_i=0} \,
\det_{\frac{w}{2}\times \frac{w}{2} }\,
\left| \Biggl( {m \atop (m/2)+\frac{l+w}{2}\, 
t_i+\lambda_j-\lambda_i }
\Biggr) \right|,
\label{combi}
\end{equation}
where $\vec t$ is a $w/2$-tupel of integers. Note that
because of the properties of the binomial coefficients,
only a finite number of terms in the sum over $\vec t$
is non-vanishing. The presence of the determinants
has to do with the fact that although all possible path
configurations appear in (\ref{combi}), the
contributions from configurations with touching or 
intersecting walkers cancel appropriately by virtue of
the alternating signs in the determinantal sum, in such
a way that only the non-intersecting ones are left over.
The largest term that can appear in any of the determinants
on the right-hand side of (\ref{combi}) is
always the product of the elements on the diagonal, namely,
\begin{equation}
\Biggl( {m \atop m/2 } \Biggr)^{\frac{w}{2}},
\label{freewalk}
\end{equation}
corresponding to the independent product of $w/2$ periodic,
free random walks consisting of $m$ steps.

\section{Adding colour and estimating the entropy}\label{colour}

The hexagon model possesses a feature that will simplify
the (asymptotic) analysis of the propagator. 
This has to do with the fact that each intersection pattern
characterizing a sandwich geometry can be brought to a
{\it standard form} by using a sequence of moves which 
affect neither the in- and outgoing 2d torus geometries nor
the value of the action (\ref{theaction}).

\begin{figure}[h]
\centerline{\scalebox{0.5}{\rotatebox{0}
{\includegraphics{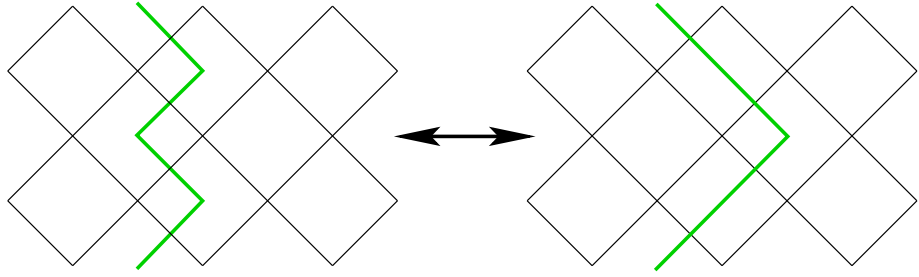}}}}
\caption[flip1]{The effect of a wedge flip on a B-C-path
drawn on the dual A-lattice.}
\label{flip1}
\end{figure}

\begin{figure}[h]
\centerline{\scalebox{0.3}{\rotatebox{0}
{\includegraphics{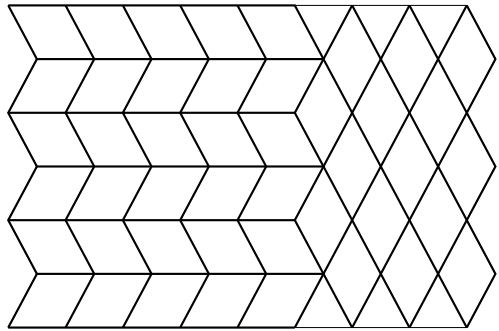}}}}
\caption[standard]{A rhombic tiling of standard form.}
\label{standard}
\end{figure}

The basic idea is to move all B-C-paths to the far left of
the strip. The elementary move necessary to achieve this
is the flip of a left-right wedge (an adjacent pair of a
dual B- and C-link) to a right-left wedge or vice versa 
(Fig.\ref{flip1}). Since any B-C-chain has by assumption an equal 
number of B- and C-links, it can be moved to the left of
the strip (where it assumes the form of a zigzag path) by a
finite sequence of such moves. In the process, it will
cross (pieces of) A-chains, but not other B-C-chains.
A typical final result after applying this procedure to all
B-C-chains of an intersection pattern is illustrated in
Fig.\ref{standard}.

\begin{figure}[h]
\centerline{\scalebox{0.5}{\rotatebox{0}
{\includegraphics{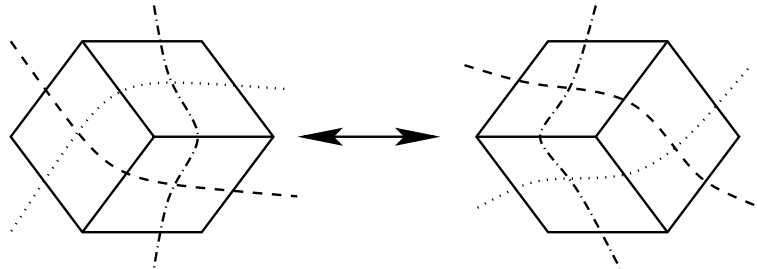}}}}
\caption[flip2]{The rearrangement of the rhombi making up
a fundamental hexagon region during
a wedge flip and of the associated dual links or lines.}
\label{flip2}
\end{figure}

In order to understand which consequences the wedge flip has
for the geometry, let us analyze which action it translates
to on the {\it unreduced} rhombic tiling. The region
on the original triangular lattice which is affected by the
wedge flip is confined to a set of six triangles forming a
single hexagon, with six dual links emerging from the
sides of the hexagon. There are two ways of tiling this
fundamental hexagon by three rhombi. In both cases,
there is a piece of B-C chain entering at the bottom of the
hexagon and coming out at the top, a piece of a
right-moving A-chain entering at the bottom left and
exiting at the top right side of the hexagon and a piece
of left-moving A-chain entering at the bottom right and
coming out at the top left. The wedge flip changes the
tiling of the fundamental hexagon and the intersection
patterns of the dual chains in the interior of the hexagon,
without altering the dual links emanating from it (Fig.\ref{flip2}).

To determine the effect on the geometry,
one has to consider all possible colourings of the
chains passing through the hexagon and how the wedge
flip affects the blue and red dual link patterns. The first
case is that where all lines have the same colour, say, blue,
which implies that the hexagon represents a gluing of
six 3-1 tetrahedra. The wedge flip only alters the way in
which we mentally divide this set of six tetrahedra into three
double-tetrahedra. Obviously this move does not in any way
affect the three-geometry, and the geometries before and
after the move should not be counted as distinct. 
However, for all
other colour choices (there is a total of six) 
the three-geometry genuinely
changes. As one can easily verify, the wedge flip in 
all of these cases corresponds
to pulling a piece of a red chain (without vertices) across
a blue-blue intersection or the other way round. This means
that the individual red and blue dual graphs are 
squeezed and stretched in the process, but remain otherwise
completely unaffected, and hence will correspond to the
same two-geometries. 

Since moreover the number of building blocks of a certain
type (3-1, 1-3 or 2-2) does not change under a flip move,
we have proved our original assertion that the wedge flip
leaves all two- and three-volumes (and hence the action) 
as well as the Teichm\"uller parameters of the two-geometries 
invariant.\footnote{
The wedge flip is an obvious candidate for a Monte-Carlo 
move in numerical simulations of the hexagon model;
it will have to be augmented by moves that
{\it can} change the $\tau$'s and other physical variables.}
We recognize here a simplified feature of the hexagon model,
compared with the most general dynamically triangulated
3d gravity model \cite{d3d4,ajl2}, even if we restricted its 
integer-$t$ slices to be flat tori. 
Namely, although the toroidal two-geometries forming
the spatial boundaries of a space-time sandwich
$\Delta t=1$ by no means fix the three-geometry
inbetween the two slices, they determine essentially uniquely
the value of the sandwich action $S(\Delta t=1)$. 
In other words, there is a large number of interpolating
space-time geometries for given, fixed boundaries, but
they all contribute with the same weight e$^{iS}$. 
{\it A main task in solving the model is therefore the computation of
the number of distinct interpolating 3-geometries between two 
adjacent flat two-tori.}

Before we can give a precise definition of this combinatorial
problem, we still need to specify how we are going to parametrize
the {\it colouring} of the rhombic intersection pattern. 
Since we have already shown that any intersection pattern
characterizing a three-geometry can
be brought to standard form without affecting the individual
red and blue torus geometries, we may without loss of
generality think of the latter as subgraphs of this standard form.
Using the notation introduced above,
we will label the uncoloured standard form
by three even integers $(w,l,m)$, where $w/2$ is the number
of zigzag B-C-chains on the left (whose individual height is
$m$) and $(l,m)$ is the size of the regular lattice of
A-rhombi on the right. 

Let us now introduce a colouring by
drawing closed dual lines onto this standard form, 
thereby producing a {\it coloured standard form} $\cal S$. There
are obviously $w/2$ independent vertical lines that can
be drawn onto the B-C-columns. We will split them
into $w_1/2$ blue lines and $w_2/2\equiv (w-w_1)/2$
red lines. It is clear that the order in which we colour these
strands will affect the three-geometry, but not the
individual two-geometries or the action. This results in a
multiplicity $\Bigl({w\atop w_1}\Bigr)$ for given in-
and out-states, counting the number of possible
orderings of blue and red vertical dual lines.

We turn next to the colouring of the remaining dual
lines, i.e. those that traverse the B-C-chains horizontally
and the A-rhombi diagonally. The choice is restricted
by the fact that the number of such lines which
are closed (and therefore can be coloured independently)
is exactly $d$ for the right-moving and $d$ for the left-moving
lines. We will denote the numbers of blue right-moving and
left-moving dual lines by $a_r$ and $a_l$, and those of
the corresponding red lines by $b_r\equiv d-a_r$ and
$b_l\equiv d-a_l$. As before, the relative ordering of
the blue and red lines in either direction leads in general
to different three-geometries, but leaves the two-geometries
and the action unchanged, thus contributing a factor
$\Bigl( {d\atop a_r}\Bigr) \Bigl( {d\atop a_l}\Bigr) $ 
to the number of interpolating states.

Putting all of these observations together, we can now 
rewrite the one-step propagator (\ref{onestep}) 
in a more explicit form. Using the essentially unique
association  $(g_1,g_2)\leftrightarrow (l,m,w_1,w_2,a_l,a_r)$ 
(c.f. Sec.\ \ref{teich} and Appendix 1), $G$ now
takes the form
\begin{equation}
G(g_1,g_2,\Delta t\! =\! 1)= 
\sum_{\cal O}
\frac{1}{C({\cal S})}\tilde M({\cal S}) 
\Bigl({w_1+w_2\atop w_1}\Bigr) \Bigl( {d\atop a_r}\Bigr) 
\Bigl( {d\atop a_l}\Bigr)\;
\e^{-S^{\rm eu}({\cal S})} .\label{ampl}
\end{equation}
The number $\tilde M({\cal S})$ counts the
distinct strip configurations that can be obtained by applying
elementary wedge flips to the coloured standard form
${\cal S}$, uniquely described by the
six parameters $(l,m,w_1,w_2,a_l,a_r)$, and the
relative order $\cal O$ of their dual coloured lines. 
(We are regarding configurations that differ by
overall translations in the $l$- and $m$-direction as equivalent.
At any rate, this choice does not affect the remainder of our
discussion.) 

In view of the discussion in Sec.\ \ref{mot}, we are interested in
the continuum behaviour of (\ref{ampl}), and in particular how
the entropy contributions compete with the kinematical action
$S^{\rm eu}$ to result in an effective (continuum) action. 
The combinatorial part $\tilde M({\cal S})$ 
of (\ref{ampl}) is very similar to that of the uncoloured
problem described in connection with Fig.\ref{walkers}, but the
colour-dependence {\it does} now enter in a slightly subtle way. 
This again has to do with the slight overcounting present in
our model (the subdivision of fundamental hexagon regions
of one colour) which implies that not all vicious-walker configurations
will correspond to distinct three-geometries. How often this occurs
depends on both the boundary geometries $g_i$ and the
relative order $\cal O$ of coloured lines of an individual
sandwich geometry. It is clear that the overcounting will be 
most pronounced when the intersection pattern has very
many dual links of one colour and very few of the other,
because this will result in many local fundamental hexagon
regions of one colour which are insensitive to wedge flips,
cf. Fig.\ \ref{flip2}. 

Let us proceed on the assumption that -- at least to leading
order -- the scaling behaviour of the entropy will not be affected
by this overcounting. This is in part justified by the numerical
investigations of \cite{ajl2}, where we found that in the
continuum limit, neighbouring spatial slices are strongly 
coupled, in the sense of 
having a similar total volume. Under this assumption,
we can drop the sum over $\cal O$ in (\ref{ampl}), and
substitute the combinatorial factor by $\tilde M=M$, where 
\begin{equation}
M(m,l,w):=\sum_{\vec\lambda}b(\vec\lambda,m,l,w)
\label{combisum}
\end{equation}
is the sum over all ordered $w/2$-tupels of initial conditions 
$\vec\lambda$
for a set of random walkers. We would like to establish the
behaviour of $M$ in the limit as $(m,l,w)$ simultaneously
become large\footnote{A natural canonical scaling ansatz is
$m,l,w \rightarrow\infty$ while sending the cutoff $a\rightarrow 0$,
in such a way that the dimensionful length variables 
$M:=am$, $L:=a l$ and $W:=a w$ remain fixed and finite.}. 
This is not completely straightforward, since 
according to (\ref{combi}) each $b$ is a sum of terms that can be
both positive and negative. 

For the purposes of this paper we will concentrate on establishing
the leading divergent behaviour of $M$; we hope to return
to the full analytic solution of the model in the near future. 
Is it the case that the leading behaviour is given by 
exp($c$ spatial volume), for some
positive constant $c$? As explained earlier, this is
needed for obtaining truly extended geometries
in the continuum limit and a necessary prerequisite for a
conformal-factor cancellation. The continuum limit that has
been considered in most of the literature on vicious walkers
is that in which both the width and the length of the lattice
become large, but {\it not} the number of walkers.
In this case one typically finds for the number $\cal N$ of
walker configurations
\begin{equation}
{\cal N}=2^{{\rm number\;
of\; walkers\; }\times {\rm number\; of\; steps}} \times
\, {\rm (subleading\; terms)}.
\label{diluted}
\end{equation}
This is precisely the type of scaling behaviour we are looking
for, but not the physical situation we are interested in,
since the paths here are ``diluted'' (even though their initial and
final points may lie close together). 

A scenario of more immediate interest to us is again that 
considered by Forrester we already cited earlier \cite{forr}, who 
treats the case of $N$ equally spaced walkers on a lattice of 
width $\mu$ with initial conditions $\lambda_i =i\nu$, $i=1,...,N$,
so that $\nu=\mu/N$ is the average distance between the walkers'
paths. However, it should be noted that his class of path
configurations is larger than ours: although he also considers
periodic boundary conditions in the $m$-direction, an individual
walker's path is not required to close on itself after $m$
steps, but may wind around the lattice several times before
doing so. The resulting closed paths will in general have
$(l,m)$-winding numbers different from (0,1) (an example is
shown in Fig.\ \ref{forbidden}) which by definition we have 
excluded from our current model. 
The evaluation of the combinatorics of this
more general case turns out to be easier because it involves
determinants of {\it cyclic} matrices, which can be
simplified. For even $m,\nu$ and odd $N$ the number of
vicious-walker configurations is given by \cite{forr}
\begin{equation}
{\cal N}^F (N,m,\nu)=\prod_{p=0}^{N-1}\, \frac{2}{\nu}\,
\sum_{b=0}^{\nu/2 -1} 2^m \cos^m (\frac{2\pi}{\nu} (\frac{p}{N} +b)).
\label{fwalks}
\end{equation}

However, it is easy to see that the overcounting involved
in (\ref{fwalks}) does not alter the leading behaviour for
large $N$ and $m$ compared to our more restricted
case. Namely, we can divide the path configurations counted
by (\ref{fwalks}) into different sectors, depending
on the lateral displacement $\Delta=\lambda_{final}-
\lambda_{initial}$ of the initial and final point of
each walker. We have $\Delta_{max}=\pm m$, with the
largest contribution to ${\cal N}^F$ coming from configurations
with $\Delta =0$ (corresponding exactly to the path
configurations with winding number 1 we are
counting in the hexagon model). The asymptotics of (\ref{fwalks})
is much easier to handle, since the right-hand side is
a product of sums over {\it positive} terms. 
Assuming an appropriate monotonic behaviour of ${\cal N}^F$,
it follows from the results in
Appendix 2 that the leading divergence for $N,m\rightarrow
\infty$ and fixed $\nu >2$ is given by
\begin{equation}
{\cal N}^F (N,m,\nu)\sim c(\nu)^{mN},\;\;\;  
2\ \cos (\frac{\pi}{\nu})\leq c(\nu ) \leq 2.
\label{fasym}
\end{equation}

Coming back to the combinatorics of the function $M$, note 
that for a given lattice width and number of walkers there are
$\Bigl( {(l+w)/2-1\atop w/2 }\Bigr) $ terms contributing
to the sum (\ref{combisum}), corresponding to all possible
initial conditions for the walkers. Since this number grows
at most exponentially in length (as opposed to volume),
the leading exponential 
behaviour will coincide with that of the largest
term in the sum, which is roughly speaking that of the
configuration(s) where the walkers are maximally distributed 
(i.e. spaced out in the $m$-direction) 
over the available space and therefore get least in the way
of each other. We can identify their average distance with
the spacing $\nu$ appearing in Forrester's formulas. (Strictly
speaking, $\nu$ is of course an (even) integer, but we do not
expect this to play a role in the continuum limit.)
We may therefore conclude that for an
average spacing $\nu\equiv (l+w)/w$, and in the limit as
$m,l,w\rightarrow\infty$, the number of 
vicious-walker configurations behaves to leading order like
\begin{equation}
M(m,l,w)\sim c(\nu)^{mw/2}\times\;{\rm (subleading\; terms)},
\label{finasym}
\end{equation}
where $c(\nu)$ is a constant of order 2, which appeared in
(\ref{fasym}).
This shows that the hexagon model of three-dimensional
quantum gravity has indeed
enough entropy, compared with more restricted cosmological
models that were studied previously.

\section{Teichm\"uller parameters for triangulated tori}\label{teich}

Our next task will be to relate the parameters 
$(l,m,w_1,w_2,a_l,a_r)$, introduced in the previous
section to label a standard coloured three-geometry, to
the variables describing the two spatial boundaries of that geometry.
In this way we will achieve a clean separation of the
geometric in- and out-data $g_1$ and $g_2$ appearing
as arguments of the amplitude $G$, and make contact
with the standard parametrization of two-dimensional flat tori. 

It is a well-known fact that in the continuum any flat torus 
geometry $g$ can be characterized 
by three numbers $(v,\tau_1,\tau_2)$, where $v$ is the
two-volume (in our case proportional to the number of triangles 
that make up the torus) and the $\tau_j$ are the two real 
Teichm\"uller parameters.\footnote{We will keep
our options open about whether we eventually want to use these or
rather the so-called moduli parameters, which label
equivalence classes of Teichm\"uller parameters with
respect to the action of the mapping class group.}
We would like to compute the geometric data 
$(v^{(i)},\tau_1^{(i)},\tau_2^{(i)})$, $i=1,2$, for the individual 
two-tori from the parameters $(l,m,w_1,w_2,a_l,a_r)$.

Recall that the numbers
of blue, red and blue-red rhombi at time $t+1/2$ is given
by $N_{31}/2$, $N_{13}/2$ and $N_{22}$ respectively. The
first two numbers give us directly the discrete two-volumes of the
spatial slices at times $t$ and $t+1$, namely, 
$v^{(1)}=N_{31}$ and $v^{(2)}=N_{13}$. 
(In order to obtain the volume in terms of the lattice
spacing $a$, one needs to multiply with the volume of
a Euclidean triangle, which is $a^2 \sqrt{3}/4$.)
Taking into account the considerations of Sec.\ \ref{comb}
on the division of the A-lattice into fundamental regions,
one derives for these numbers the expressions
\begin{eqnarray}
N_{31}&=&w_1 \frac{m}{2d}(a_r+a_l)+
\frac{lm}{d^2} a_r a_l,\nonumber\\
N_{13}&=&w_2 \frac{m}{2d}(b_r+b_l)+
\frac{lm}{d^2} b_r b_l,\nonumber\\
N_{22}&=&\frac{m}{4d}(w_1 (b_r+b_l) +w_2 (a_r+a_l))+
\frac{lm}{2d^2}(a_r b_l+a_l b_r).
\label{volumes}
\end{eqnarray}

In order to determine the Teichm\"uller parameters for the
tori, we will identify for each colour (blue or red) a pair
of oriented, closed geodesics, i.e. straight lines of
minimal length on the 
corresponding torus at time $t$ or $t+1$. (Note that these
do not have to coincide with any triangle edges.) 
Each pair consists of
a circle with winding number $(1,0)$ and one with
winding number $(0,1)$ 
when viewed as curves on the intersection pattern at
$t+1/2$ (the two winding numbers refer to the horizontal 
$l,w$-direction and the vertical $m$-direction).
The winding numbers of closed curves in the spatial
two-geometries can therefore be thought of as
``inherited'' from the three-dimensional geometry.
These curves are clearly unique up to two-dimensional translations 
(if we think of the flat tori at integer time $t$ as being
rolled out into the two-dimensional plane). For a given colour,
their lengths and their relative intersection angle
are translation-invariant, and in one-to-one
correspondence with the torus parameters $(v,\tau_1,\tau_2)$.

\begin{figure}[h]
\centerline{\scalebox{0.5}{\rotatebox{0}
{\includegraphics{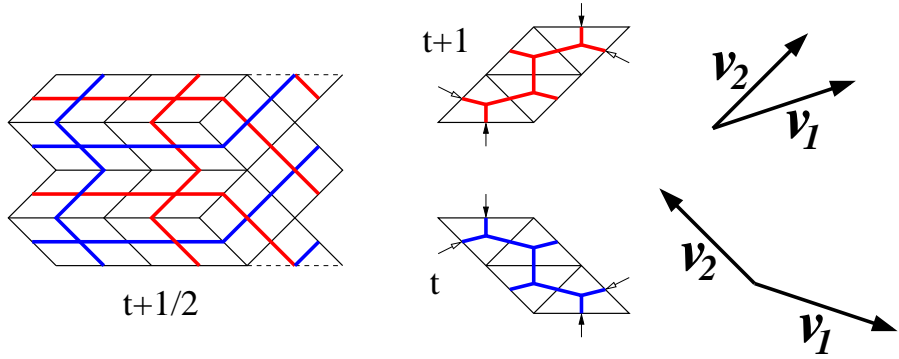}}}}
\caption[vectors]{An intersection pattern at time $t+1/2$ gives
rise to a blue triangulation at time $t$ and a red triangulation
at time $t+1$. We also show the two vectors ${\cal V}_1$ and
${\cal V}_2$ representing the
closed geodesics of winding number (1,0) and (0,1) for both
of these geometries.}
\label{vectors}
\end{figure}

Let us look at a simple example in order to illustrate this
procedure. Fig.\ref{vectors} shows on the left 
a coloured intersection
pattern in standard form representing a sandwich geometry, 
and consisting of 12 rhombi. It gives rise to two spatial
boundary geometries, both of them flat tori with four
triangles each. They have been cut open and are represented
by two parallelograms in the plane, where the small black and
white arrows 
indicate how their opposite sides are to be
identified pairwise. Each of these two-geometries has oriented
geodesics with winding number (1,0) and (0,1) respectively.
They can be represented by a pair of vectors 
$({\cal V}_1,{\cal V}_2)$ drawn onto the parallelogram.
For ease of representation, the vectors are depicted 
on the right with
their correct lengths and orientations,
and with a common origin. 

It remains to compute the lengths and angles of these
vectors from
the data $(l,m,w_1,w_2,a_r, a_l, b_r,b_l)$ characterizing
the coloured intersection pattern in standard form. 
Representing the blue closed geodesic with winding number $(1,0)$ 
at time $t$ as a two-dimensional vector ${\cal V}_1^{(1)}$ and 
the blue closed geodesic with winding number $(0,1)$ as a second
vector ${\cal V}_2^{(1)}$, one finds in terms of the discrete
units inherited from the $(l,m)$-coordinate system at
$t+1/2$  
\begin{eqnarray}
{\cal V}_1^{(1)} &=&\Bigl( \frac{l}{2d}(a_l+a_r)+w_1,
\frac{l}{2d}(a_l-a_r)\Bigr),\nonumber\\
{\cal V}_2^{(1)} &=&\Bigl(\frac{m}{2d}(a_l-a_r),
\frac{m}{2d}(a_l+a_r)\Bigr) \label{vec1}.
\end{eqnarray}
Translating this into absolute units, one finds
\begin{eqnarray}
{\cal V}_1'^{(1)} &=&\Bigl( \frac{l}{4d}(a_l+a_r)+\frac{w_1}{2},
\frac{\sqrt{3}l}{4d}(a_l-a_r)\Bigr) a,\nonumber\\
{\cal V}_2'^{(1)} &=&\Bigl( \frac{m}{4d}(a_l-a_r),
\frac{\sqrt{3}m}{4d}(a_l+a_r)\Bigr) a\label{vec2},
\end{eqnarray}
where $a$ again denotes the lattice spacing.
The two vectors ${\cal V}_1'^{(2)}$, ${\cal V}_2'^{(2)}$
characterizing the red triangulation
at time $t+1$ are obtained by substituting
$a_l,a_r,w_1$ in (\ref{vec2}) by $b_l,b_r,w_2$. 
Note that the relations (\ref{vec2}) imply the inequalities
\begin{equation}
{\cal V}_{1,1}'^{(i)}\geq 0, \;\; 
{\cal V}_{2,2}'^{(i)}\geq 0, \;\;
{\cal V}_{1,1}'^{(i)}\geq |{\cal V}_{1,2}'^{(i)}|, \;\;
{\cal V}_{2,2}'^{(i)}\geq |{\cal V}_{2,1}'^{(i)}|.
\label{vinequalities}
\end{equation}
As is shown in Appendix 1, the coordinate transformation
between the six variables $(l,m,w_1,w_2,a_l/d,a_r/d)$ and
the components of the vectors ${\cal V}_j'^{(i)}$,
subject to the constraints
\begin{equation}
{\cal V}_{1,2}'^{(2)} =-{\cal V}_{1,2}'^{(1)},\;\;
{\cal V}_{2,1}'^{(2)} =-{\cal V}_{2,1}'^{(1)},
\label{vconstraints}
\end{equation}
is one-to-one.

\begin{figure}[h]
\centerline{\scalebox{0.5}{\rotatebox{0}
{\includegraphics{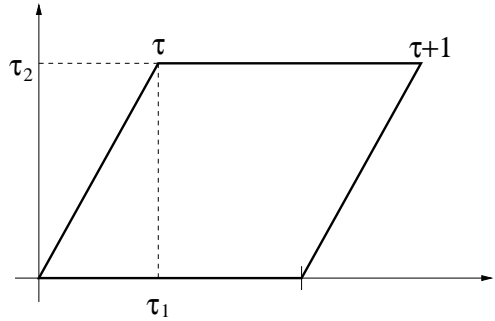}}}}
\caption[tauplane]{Standard representation of a flat 
torus with Teichm\"uller parameter $\tau =\tau_1+i \tau_2$ 
as a parallelogram in the complex $\tau$-plane.}
\label{tauplane}
\end{figure}

Up to a common rotation and a global rescaling, 
each pair of vectors $({\cal V}_1'^{(i)},{\cal V}_2'^{(i)})$ 
can be identified with the
vectors spanning the parallelogram in the
standard representation of a flat torus with
normalized area in the complex $\tau^{(i)}$-plane
($\tau^{(i)}=\tau_1^{(i)} +i\tau_2^{(i)}$), Fig.\ref{tauplane}. 
It is straightforward to compute the SO(2)-rotation 
\begin{equation}
\Bigl(\matrix{\;\; \cos\phi^{(i)} & \sin\phi^{(i)} \cr 
 -\sin\phi^{(i)} & \cos\phi^{(i)}}\Bigr)
\label{rot}
\end{equation}
which aligns the vector ${\cal V}_1'^{(i)}$ with the 
$\tau_1^{(i)}$-axis. One finds
\begin{equation}
\cos\phi^{(1)} =\frac{a}{L^{(1)}}(\frac{l}{4d}(a_l+a_r)+\frac{w_1}{2}),
\hspace{1cm} \sin\phi^{(1)} =\frac{a}{L^{(1)}}\frac{\sqrt{3}l}{4d} 
(a_l-a_r),
\label{angles}
\end{equation}
where $L^{(1)}$ denotes the length of the vector ${\cal V}_1'^{(1)} $,
\begin{equation}
L^{(1)}:=||{\cal V}_1'^{(1)}||=
\sqrt{\Bigl(\frac{l}{4d}(a_l+a_r)+\frac{w_1}{2}\Bigr)^2+
\Bigl(\frac{\sqrt{3}l}{4d}(a_l-a_r)\Bigr)^2}\, a.
\label{length}
\end{equation}
The corresponding expressions for $\cos\phi^{(2)}$, 
$\sin\phi^{(2)}$, $L^{(2)}$
are obtained by substituting $(a_l,a_r,w_1)\mapsto (b_l,b_r,w_2)$.

We still have to rescale all lengths by a factor $1/L^{(1)}$
so that the rotated vector ${\cal V}_1'^{(1)}$ assumes
length 1. Applying then both the rotation 
(\ref{rot}), (\ref{angles})
and this rescaling to the vector ${\cal V}_2'^{(1)}$, we can
read off directly the dimensionless 
Teichm\"uller parameters $\tau_i^{(1)}$.
Collecting the geometric data describing uniquely the
blue torus at time $t$, we have finally
\begin{eqnarray}
v^{(1)}&=&
\Bigl( a_l a_r \frac{lm}{d^2} +
(a_l+a_r)\frac{mw_1}{2d}\Bigr),\nonumber\\
\tau_1^{(1)}&=&\frac{m}{4d(L^{(1)})^2}\Bigl( (a_l^2-a_r^2)\frac{l}{d}+
(a_l-a_r)\frac{w_1}{2}\Bigr) a^2,\nonumber\\
\tau_2^{(1)}&=&\frac{m}{4d(L^{(1)})^2}\Bigl(a_l a_r\frac{\sqrt{3}l}{d}+
(a_l+a_r)\frac{\sqrt{3}w_1}{2}\Bigr) a^2\equiv 
\frac{\sqrt{3}}{4} \frac{a^2}{(L^{(1)})^2} v^{(1)}.
\label{torusdata}
\end{eqnarray}
The map between the independent vector components 
${\cal V}_{i,j}'^{(k)}$ (or, equivalently, the sandwich
variables $(l,m,w_1,w_2,a_l/d,a_r/d)$) and the
$(v^{(k)},\tau^{(k)})$ is in general two-to-one
(see Appendix 1).

One may also wish to reexpress the action in terms of these
torus parameters. Because the transformation to these
parameters is not bijective, this can only be done 
modulo a sign ambiguity.
One first writes the
counting variables $N_{ij}$ as functions of the two-vectors
${\cal V}'^{(k)}$ and then in turn expresses the latter as
functions of the torus parameters. Details of these calculations
can again be found in Appendix 1. The explicit form for the
sandwich action $S^{\rm eu}({\cal S})=S^{\rm eu}(v^{(1)},\tau^{(1)},
v^{(2)},\tau^{(2)})$ is given by (\ref{theaction}), with
\begin{eqnarray}
N_{31}&\! \!=\!\! &  v^{(1)}\nonumber\\
N_{13}&\! \!=\!\! & v^{(2)}\nonumber\\
N_{22}&\! \!=\!\! &\frac{1}{2} \frac{1}{(\tau_1^{(1)}+\tau_1^{(2)})^2 +
(\tau_2^{(1)}-\tau_2^{(2)})^2} \Biggl( -2 (\tau_1^{(1)} \tau_2^{(2)} +
\tau_1^{(2)} \tau_2^{(1)}) \Bigl( v^{(1)}\frac{ \tau_1^{(1)}}{\tau_2^{(1)} }+
v^{(2)}\frac{ \tau_1^{(2)}}{\tau_2^{(2)} } \Bigr) \nonumber\\
&-&\Omega\; \Bigl( (\tau_1^{(1)})^2 -(\tau_1^{(2)})^2 +(\tau_2^{(1)})^2 -
(\tau_2^{(2)})^2 \Bigr) \Biggr).
\label{long}
\end{eqnarray}
The function 
\begin{equation}
\Omega = 
({\cal V}_{1,1}'^{(1)} {\cal V}_{2,2}'^{(2)}-
{\cal V}_{2,2}'^{(1)} {\cal V}_{1,1}'^{(2)})\frac{4}{a^2 \sqrt{3}}
\label{omega1}
\end{equation}
can be expressed in terms of the $(v^{(k)},\tau^{(k)})$
only up to a sign, namely, as the positive and negative
square-root of
\begin{equation}
\Omega^2 = \frac{ v^{(1)} v^{(2)} }{ \tau_2^{(1)}\tau_2^{(2)} } 
\Bigl( (\tau_1^{(1)}+\tau_1^{(2)})^2 +
(\tau_2^{(1)}-\tau_2^{(2)})^2 \Bigr) -
\Bigl( v^{(1)}\frac{ \tau_1^{(1)}}{\tau_2^{(1)} }+
v^{(2)}\frac{ \tau_1^{(2)}}{\tau_2^{(2)} } \Bigr)^2.
\label{omega2}
\end{equation}

The resulting action has a feature that may at first seem
puzzling. Let us think of the exponentiated action for fixed
boundary geometries as a matrix element (c.f. equation 
(\ref{trans})),
\begin{equation}
{\rm e}^{-S^{\rm eu}(v^{(1)},\tau^{(1)},
v^{(2)},\tau^{(2)})} =:
\langle v^{(2)},\tau^{(2)}|\hat t | v^{(1)},
\tau^{(1)}\rangle.
\label{melement} 
\end{equation}
Usually this kinematical ``transfer matrix'' can be written as
$\hat t= {\rm e}^{-a\hat h}={\bf 1} -a \hat h +O(a^2)$, 
where $\hat h$ denotes the discrete, kinematical Hamilton
operator of the system\footnote{This is the ``kinematical'',
and not yet the full, ``effective'' Hamiltonian, because
we have not included any entropy contributions in the matrix
element.}, and where we have reintroduced the discrete
unit $\Delta t=a$ for a single time step. 

For such an expansion to exist, in the
limit of small $\Delta t$ the configuration variables at
a neighbouring spatial slice should always be expressible as
$v(t+a)\simeq v(t)+a \dot v$, $\tau (t+a)\simeq \tau (t)+a\dot \tau$.
However, looking at the explicit formulas for $\tau_1^{(1)}$
and $\tau_1^{(2)}$ (representing the variable $\tau_1$ at two
adjacent spatial slices) as functions of $a_l$ and $a_r$, one sees 
that they always have opposite signs. It is therefore
kinematically impossible to set them equal
{\it unless} they both vanish (or, equivalently, $a_l=a_r$).
This can be traced back to the fact that a natural ``pairing''
between two neighbouring spatial geometries 
occurs in our construction
by putting their two graphs together in an intersection
pattern. As can be seen from the elementary example
depicted in Fig.\ref{vectors}, the natural ``dual'' of a
particular blue graph is given not by the graph itself, but
by its reflection, where the relative orientation of
the two coordinate axes has been reversed. In terms of the
Teichm\"uller parameters (since by definition we always keep
$\tau_2 \geq 0$) this amounts to a map $R:\tau_1\mapsto -\tau_1$,
with a lowest-order expansion $\hat t =R +O(a)$ for the
matrix $\hat t$.
From this point of view, a more natural elementary time unit
in our model consists of two time steps, a feature that has
also been observed in other discrete models of 2+1 gravity
\cite{networks}.

\section{A discrete sampling of the Teichm\"uller and moduli spaces}
\label{sample}

Although we have established above a description of the geometry
of spatial slices in terms of the standard Teichm\"uller parameters,
it is clear that in our model the parametrization will not be a
continuous one, since the $\tau^{(i)}$ according to (\ref{torusdata})
are given as functions of
certain discrete (integer) parameters characterizing triangulated
geometries.  From a continuum point of view our simplicial
discretization therefore provides a discrete sampling of the Teichm\"uller
space $\cal T$ (which topologically is an ${\bf R}^2$).  
It is interesting to look in more detail at how this sampling works
as function of a given cutoff on the volume of either the space-time 
sandwiches or the spatial slices, say. 

In addition, one might be interested in the configuration space
obtained from Teich\-m\"uller space by identifying points that
differ by the action of  ``large diffeomorphisms'' (the
so-called moduli space), i.e.
those two-dimensional transformations that do not lie
in the connected component of the diffeomorphism
group of the torus. There are different suggestions of how
these transformations should be incorporated into the quantum 
theory of 3d gravity, whether as exact invariances, as 
symmetries with a unitary action, or possibly not at all (see 
\cite{twists} for further discussions). We will not take any
particular viewpoint here, apart from remarking that 
quotienting out by large diffeomorphisms usually leads to 
additional complications in the quantum theory . 

In order to determine the modulus of a given Teichm\"uller parameter
$\tau =\tau_1 +i \tau_2$, we have to map it by a sequence
of modular transformations
\begin{equation}
\tau \rightarrow \tau +1,\;\;\;\;\;\;\;\; \tau\rightarrow \frac{\tau}{\tau +1}
\label{mod}
\end{equation}
(the generators of the mapping class group of the torus) to a
point in a fundamental region $\subset\cal T$, usually taken
to be the ``keyhole region'' defined by $-1/2\leq \tau_1 \leq 1/2$ and
$| \tau |^2\equiv \tau_1^2+\tau_2^2 \geq 1$. 

To get a qualitative idea of the nature of the discrete sampling
of these two spaces, we have set up a small program that
generates all values of $(v,\tau)$ that occur for sandwich
geometries with parameters $0\leq l,m,w \leq 10$ (and with
the lattice cutoff $a$ set to 1). This is easy to implement, but 
does not quite amount to a systematic bound on the total space-time 
volume (which is proportional to $V=(l+w)m$). The range of
$V$ is between 4 and 200.
Geometries with one or two vanishing spatial volumes $v^{(i)}$ 
were not allowed. 

\begin{figure}[h]
{\center
\begin{tabular}{ccc}
\includegraphics[width=9cm, height=6cm]{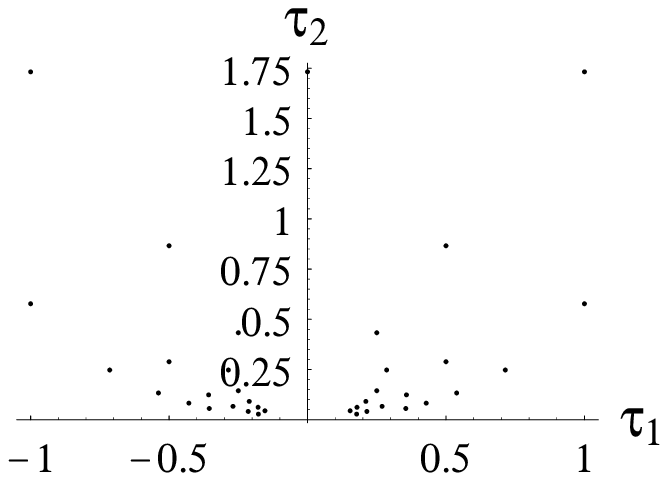}
     &  \hspace{-2cm} & 
\includegraphics[width=9cm,height=6cm]{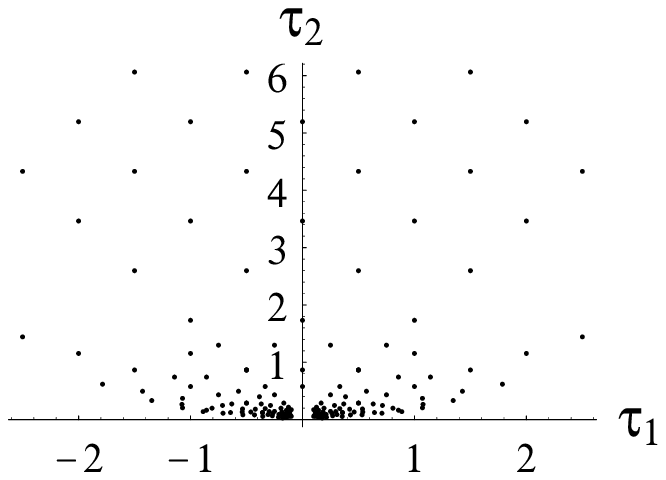}\\
\includegraphics[width=9cm, height=6cm]{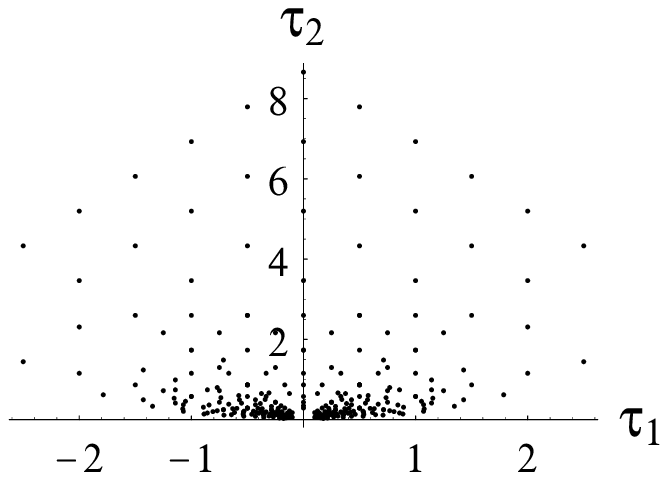}
     &  \hspace{-2cm} & 
\includegraphics[width=9cm,height=6cm]{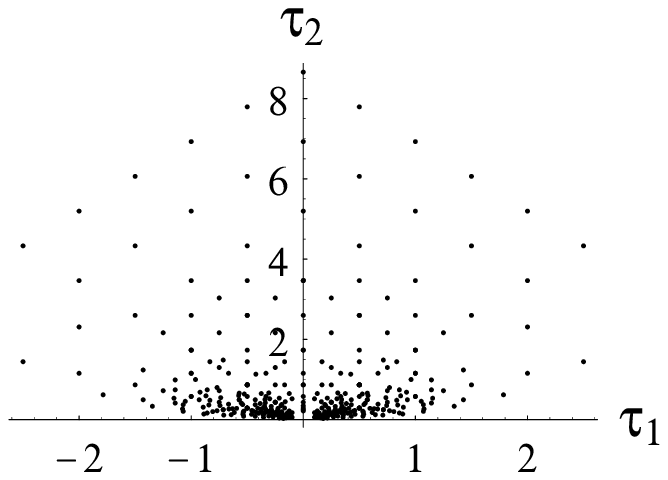}
    \end{tabular}
\par} \caption{\label{firstset}Sampling Teichm\"uller space,
for maximal spatial volumes $v_{max}=4,14$ (top row) and
24,34 (bottom row).}
\end{figure}

\begin{figure}[h]
\centerline{\scalebox{1.2}{\rotatebox{0}
{\includegraphics{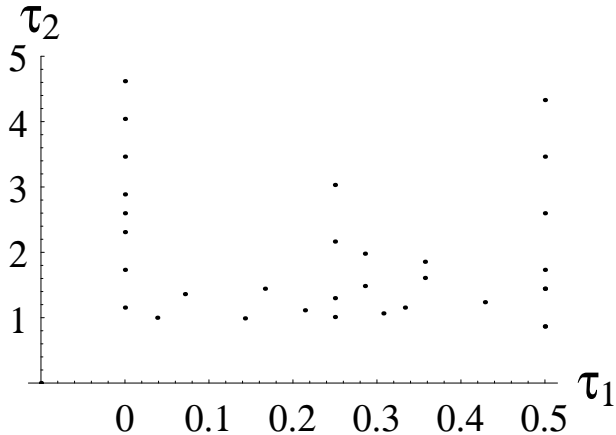}}}}
\caption[modul]{A sample of moduli parameters inside the right
half of the keyhole region.}
\label{modul}
\end{figure}

We show in Fig.\ref{firstset} a sequence of samplings of the upper
half of the complex $\tau$-plane, with dots indicating the 
$\tau$-values that
occur for a spatial slice of {\it torus} volume (the number of triangles,
which is always even) $v\leq v_{max}$,
where $v_{max}=4,14,24,34$. The origin of the Teichm\"uller
space is clearly an accumulation point, with sampled points
spreading out more and more with growing torus volume. 
The points are arranged symmetrically about the imaginary
axis and seem to lie along various well-defined curves and straight
lines through the origin. Note that any given point may occur at
more than one volume; the information about this multiplicity
is not included in our plots. 
Fig.\ref{modul} shows half of the keyhole
region with all moduli that occur up to torus volumes 32 (again,
the left half of the region can be obtained by reflection). 
Comparing with the number of different points appearing 
in Fig.\ref{firstset}, it is clear that the modular parameters are
highly degenerate. -- It is clearly of interest to study
these distributions more systematically and for larger volumes,
and compare them with the natural measures on Teichm\"uller
and moduli space, but this would lead us beyond the scope
of the present paper.

\section{Conclusions}\label{concl}

We have introduced a dynamically triangulated model of
three-dimensional Lorentzian quantum gravity whose spatial slices
at integer times are flat two-tori. As we have shown, this symmetry 
restriction 
simplifies an exact analysis considerably: the spatial slices
(and therefore the associated states of the Hilbert space) are
labelled by just three parameters -- two Teichm\"uller parameters
and a global ``conformal factor" -- and the evaluation of the model's
entropy is related to that of a set of vicious walkers, which is a rather
well-studied combinatorial problem. 

It should be pointed out that our model is not a 
``minisuperspace model'' in the usual sense of the word
(``reduce classically and then quantize''),
since the spatial slices at non-integer $t$ are in general
not translationally invariant. Therefore, more geometric degrees 
of freedom contribute to the superposition of space-time histories 
in the path integral than is obvious from the integer-$t$ slices. 
(In this sense it is related in spirit to recently studied 
cosmological models in canonical loop quantum gravity, where
also part of the reduction occurs only at the quantum level,
see, for example, \cite{bojo}.) As we already explained in
Sec.\ \ref{mot}, the conditions on the {\it space-time} 
geometries cannot be too stringent in dynamically triangulated
formulations of quantum gravity, in order to have a
sufficiently large entropy and a potentially interesting continuum
limit\footnote{One could argue that this is actually desirable from 
a physical
point of view. The more restrictive a mini-/midi-superspace model
is, the less likely it is that its dynamics is representative of
that of full gravity (see \cite{kuchryan} for further discussion
and an explicit example)}.

Crucially, we could show that the hexagon model does have
enough entropy in the sense that the number of triangulations
of a sandwich $\Delta t=1$ with given boundaries to leading
order scales exponentially with the volume of the slice. 
We also exhibited explicitly how the discrete triangulation data 
can be translated into the more familiar parametrization of the
flat two-tori in terms of Teichm\"uller (or moduli) variables.

All of these properties give rise to the hope that the hexagon
model will provide a link between the LDT formulation of
three-dimensional quantum gravity and alternative, reduced phase 
space quantizations. On the one hand there is a good chance it
will lie in the same universality class as the full dynamically
triangulated model (after all, the degrees of freedom we got rid
of by working with flat spatial slices are those of the local
conformal mode, which are known to be unphysical), and
on the other hand the parametrization of the Hilbert space
and the quantum Hamiltonian will be close to that of
the canonical formulations. 

What remains to be analyzed is the precise nature of the
continuum limit of the hexagon model. What is the subleading
asymptotic behaviour of the state sum? Can we reconfirm our
earlier conclusion \cite{ajl2,ajl3,ajlv} that the gravitational 
constant
is not renormalized, but merely sets an overall scale?
How is the divergence coming from the (global) conformal
mode in the action compensated by a corresponding term
in the entropy? What is the functional form of the effective
continuum Hamiltonian in terms of the Teichm\"uller 
parameters? What is its groundstate and is it identical
to the groundstate seen in the numerical simulations 
reported in \cite{ajl2,ajl3,ajl4}? 

One may wonder how we can hope to make much analytical
progress in the investigation of a {\it three}-dimensional
statistical model. Firstly, the answer is of course that 
{\it pure} gravity is a very special type of theory in three
dimensions, which is known to possess only a finite number
of global metric degrees of freedom. Secondly, as we have
shown, an essential part of the combinatorics of the
hexagon model's transfer matrix is that of a 
{\it two}-dimensional problem of vicious walkers, about which 
a number of analytic results are already available. 
In summary, we think that a further investigation of this
model is a promising avenue to pursue, both to
advance our understanding of dynamically triangulated
models and their continuum limits, and to achieve
some degree of unification among the existing and rather 
disparate approaches to three-dimensional quantum gravity.

\vspace{0.7cm}
\noindent {\it Acknowledgements.} R.L. wishes to thank 
C.\ Krattenthaler for correspondence on vicious walkers, a critical
reading of the paper, and for 
pointing out reference \cite{grab}. She also thanks J.\ Ambj\o rn
for discussions and C.\ Dehne for comments on an earlier version
of this manuscript. Lastly, she acknowledges support by the EU
network on ``Discrete Random Geometry", grant 
HPRN-CT-1999-00161, and by the ESF network no. 82 on
``Geometry and Disorder".

\vspace{0.5cm}

\section{Appendix 1}

In this appendix we give some details of the coordinate
transformations used in the main text in Sec.\ref{teich}.
As a first step, we give the transformation law from
(an independent subset of) the combinatorial parameters 
describing the discretized space-time slices to the
dimensionful vectorial quantitites ${\cal V}'$. 
Taking into account the identites $w_1+w_2=w$, $a_l+b_l=d$,
$a_r+b_r =d$ (with $d:={\rm GCD}(l/2,m/2)$), we choose
as an independent set of the former the six variables
$(l,m,w_1,w_2,a_l/d,a_r/d)$. Note also that not all of the
components of the two-vectors ${\cal V}'^{(i)}$, $i=1,2$, 
are independent, but we have the two constraints
(\ref{vconstraints}).
The set of transformation laws is given by
\begin{eqnarray}
&&al=\frac{2}{3} 
\frac{{\cal V}_{1,2}'^{(1)}}{{\cal V}_{2,1}'^{(1)}}
({\cal V}_{2,2}'^{(1)}+{\cal V}_{2,2}'^{(2)}),\;\;\;\;
am=\frac{2}{\sqrt{3}} 
({\cal V}_{2,2}'^{(1)}+{\cal V}_{2,2}'^{(2)}),\;\;
\nonumber\\
&&aw_1= 2{\cal V}_{1,1}'^{(1)} -\frac{2}{3} 
\frac{{\cal V}_{1,2}'^{(1)} {\cal V}_{2,2}'^{(1)}}
{{\cal V}_{2,1}'^{(1)}},\;\;\;\; 
aw_2= 2{\cal V}_{1,1}'^{(2)} -\frac{2}{3} 
\frac{{\cal V}_{1,2}'^{(1)} {\cal V}_{2,2}'^{(2)}}
{{\cal V}_{2,1}'^{(1)}},\;\; \nonumber\\
&&\frac{a_l}{d}=\frac{\sqrt{3} {\cal V}_{2,1}'^{(1)} +
{\cal V}_{2,2}'^{(1)} }
{ {\cal V}_{2,2}'^{(1)}+{\cal V}_{2,2}'^{(2)} },\;\;\;\;
\frac{a_r}{d}=\frac{-\sqrt{3} {\cal V}_{2,1}'^{(1)} +
{\cal V}_{2,2}'^{(1)} }
{ {\cal V}_{2,2}'^{(1)}+{\cal V}_{2,2}'^{(2)} }.
\label{atov}
\end{eqnarray}
This is indeed a well-defined coordinate transformation,
with Jacobian 
\begin{equation}
\tilde J = \frac{64}{3}\frac{1}
{({\cal V}_{2,2}'^{(1)}+{\cal V}_{2,2}'^{(2)})^2}\,
\frac{ {\cal V}_{2,2}'^{(2)} }{ {\cal V}_{2,1}'^{(1)} }. 
\label{jacob2}
\end{equation}

Next, let us establish the coordinate transformation that
will enable us to rewrite part of the action in terms of two-volumes
and Teichm\"uller parameters. Note first that
\begin{eqnarray}
a^2 \frac{\sqrt{3}}{4}\ N_{31}&= &
{\cal V}_1'^{(1)}\times {\cal V}_2'^{(1)}
\nonumber\\
a^2 \frac{\sqrt{3}}{4}\ N_{13}&=&
{\cal V}_1'^{(2)}\times {\cal V}_2'^{(2)}
\nonumber\\
a^2 \frac{\sqrt{3}}{2}\ N_{22}&=&
{\cal V}_1'^{(1)}\times {\cal V}_2'^{(2)} +
{\cal V}_1'^{(2)}\times {\cal V}_2'^{(1)}.
\label{nagain}
\end{eqnarray}
These relations can be verified using the transformation
laws (\ref{atov}) from the independent variables 
$(l,m,w_1,w_2,a_l/d,a_r/d)$ to the vectorial quantities
${\cal V}'$. 

It is more involved to express the components of the
two-vectors ${\cal V}'^{(i)}_j$ in terms of
the torus parameters, subject to the constraints
(\ref{vconstraints}). We need to invert the relations
\begin{eqnarray}
\tau_1^{(i)}&=&\frac{ {\cal V}_1'^{(i)}\cdot{\cal V}_2'^{(i)}}
{|| {\cal V}_1'^{(i)} ||^2},\nonumber\\
\tau_2^{(i)}&=&\frac{ {\cal V}_1'^{(i)}\times{\cal V}_2'^{(i)}}
{|| {\cal V}_1'^{(i)} ||^2},\nonumber\\
v^{(i)}&= &
{\cal V}_1'^{(i)}\times{\cal V}_2'^{(i)}\ \frac{4}{a^2 \sqrt{3}}.
\label{tautov}
\end{eqnarray}
This gives rise to certain algebraic 
expressions for the independent
vector components $({\cal V}_{1,1}'^{(1)}, {\cal V}_{1,2}'^{(1)} , 
{\cal V}_{2,1}'^{(1)}, 
{\cal V}_{2,2}'^{(1)}, {\cal V}_{1,1}'^{(2)}, {\cal V}_{2,2}'^{(2)})$ 
in terms of the Teichm\"uller parameters and the two-volumes, 
which we do not bother to write down explicitly here. 
The Jacobian of this transformation is 
\begin{equation}
J= 4\Omega  
({\cal V}_1'^{(1)}\times {\cal V}_2'^{(1)} )
({\cal V}_1'^{(2)}\times {\cal V}_2'^{(2)} )
|| {\cal V}_1'^{(1)} ||^{-4}  || {\cal V}_1'^{(2)} ||^{-4},
\label{jacob1}
\end{equation}
where $\Omega$ was already defined earlier in (\ref{omega1}).
Substituting the vector components into the equations
(\ref{nagain}), one obtains the expressions (\ref{long})
given in the main text. The fact that this coordinate
transformation is not bijective finds its expression in 
the fact that there are two regions in ``$\cal V$-space''
where $J$ has opposite signs, and which are separated by
the hypermanifold defined by $\Omega =0$. In geometric
terms, this comes about because for two pairs of 
two-vectors $\{ {\cal V}_1^{(i)}, {\cal V}_2^{(i)} \}$,
$i=1,2$ (where for each $i$ the lengths and relative
angle of ${\cal V}_1^{(i)}$ and $ {\cal V}_2^{(i)}$
are uniquely specified by $(\tau_1^{(i)},\tau_2^{(i)},v^{(i)})$
according to (\ref{tautov})) there are in general two
ways in which these two vectors pairs can be arranged
relative to each other (for example, by specifying the
angle between ${\cal V}_1^{(1)}$ and ${\cal V}_1^{(2)}$)
which satisfy the constraints (\ref{vconstraints}) on the vector
components, as well as the inequalities (\ref{vinequalities}).
Note that this does not imply that if one of these
$\cal V$-configurations can be realized as a three-dimensional
triangulation (i.e. corresponds to suitable
discrete values of $(l,m,w_1,w_2,a_l/d,a_r/d)$) that the other one
of the pair can too. In fact, this is in general not the case.

\section{Appendix 2}

In this appendix we determine the leading asymptotic behaviour
of the combinatorics of a set of equally spaced vicious walkers
on a lattice with torus topology, as quoted in
Sec.\ \ref{colour} of the main text. Following \cite{forr},
the number of configurations of $N$ vicious walkers whose
initial and final coordinates are equally spaced at a distance
$\nu$ and who walk for $m$ steps is given by
\begin{equation}
{\cal N}^F (N,m,\nu)=\prod_{p=0}^{N-1}\, \frac{2^{m+1}}{\nu}\, 
\sum_{b=0}^{\nu/2 -1} \cos^m (\frac{2\pi}{\nu} (\frac{p}{N} +b)),
\label{fwalksagain}
\end{equation} 
where $m$ and $\nu$ are assumed even and $N$ odd. The width of
the lattice is therefore $\mu =\nu N$ and its height is $m$.
We are interested in the asymptotic behaviour of (\ref{fwalksagain})
as $m$ and $N$ become both large. Assuming that $\nu >2$, let
us for some given $p$ consider the corresponding sum over $b$ on the 
right-hand side. It is clear that for large $m$, this will
be dominated by the term with the largest absolute value of
the cosine. For $p\leq N/2$, this is the term where $b$ is
minimal, i.e. $b=0$, and for $p\geq N/2$ the term where $b$ is
maximal. We thus arrive at
\begin{eqnarray}
{\cal N}^F &\!\sim\! & 2^{mN} \prod_{p=0}^{\frac{N-1}{2}}
\cos^m (2\pi \frac{p}{\nu N}) \prod_{p=\frac{N+1}{2}}^{N-1}
\cos^m (2\pi (\frac{p}{\nu N} +\frac{1}{2} -\frac{1}{\nu}) )
\nonumber\\
&=& 2^{mN}\ \Bigl[  \prod_{p=0}^{\frac{N-1}{2}} \cos^2 
(\frac{2\pi p}{\nu N}) \Bigr]^m,
\label{asy1}
\end{eqnarray}
where we have already dropped terms that do not scale
exponentially with the volume. Next, let us make an estimate
of the product in the square bracket for large $N$.
Since for all $N$ the argument of the cosine lies in the
interval $[0,\pi/\nu ]$ and since for large $N$ we have
\begin{equation}
\cos^N\Bigl(\frac{\pi}{\nu} (1-\frac{1}{N})\Bigr) \sim
(\cos \frac{\pi}{\nu})^N,
\label{cosest}
\end{equation}
the leading asymptotic behaviour of ${\cal N}^F$ can
be estimated by
\begin{equation}
(2\ \cos\frac{\pi}{\nu})^{mN} \leq {\cal N}^F(N,m,\nu)
\leq 2^{mN},\;\; \nu >2.
\label{endest}
\end{equation}
The argument leading to (\ref{endest}) is not yet water-tight,
since we first let $m$, and only afterwards $N$ become large.
That there are no further terms contributing at the same
order can be seen by establishing an upper bound for
the potentially dangerous part of ${\cal N}^F$,
\begin{eqnarray}
\prod_{p=1}^{\frac{N-1}{2}} \, 
\Biggl[ \sum_{b=0}^{\nu/2 -1} \cos^m (\frac{2\pi}{\nu} (\frac{p}{N} +b))
\Biggr]^2  \!\!\!\! &\leq & \!\!\!\! 
\prod_{p=1}^{\frac{N-1}{2}} \, \Biggl[ \cos^m (\frac{2\pi p}{\nu N} )\,
\Bigl( 1+ (\frac{\nu}{2}-1)
\frac{ \cos^m (\frac{2\pi}{\nu} (1-\frac{p}{N})) }{\cos^m (\frac{2\pi p}{\nu N} )}
\Bigl) \Biggl]^2\nonumber \\
\leq  
\prod_{p=1}^{\frac{N-1}{2}} \, \Biggl[ \cos^m (\frac{2\pi p}{\nu N} )\,
\!\!\!\! &&\!\!\!\!\!\!\!\!\!\!\! \Bigl( 1+(\frac{\nu}{2}-1)
\frac{ \cos^m (\frac{\pi}{\nu} (1+\frac{1}{N})) }
{\cos^m (\frac{\pi }{\nu}(1-\frac{1}{N}) )}
\Bigl) \Biggl]^2\nonumber \\
= \Bigl( 1+ (\frac{\nu}{2}-1) (1\, - \!\!\!\! &&\!\!\!\!\!\!\!\!\!\!\! \frac{2\pi}{\nu N}
\tan \frac{\pi}{\nu})^m
\Bigr)^{N-1} 
\Biggl[ \prod_{p=1}^{\frac{N-1}{2}} \, \cos (\frac{2\pi p}{\nu N} )\Biggr]^{2m}.
\label{estim1}
\end{eqnarray} 
Letting now $m$ and $N$ simultaneously 
become large, one sees that the
term in front of the square bracket in the last line of (\ref{estim1})
scales at most exponentially with $m$, and not with the volume
$mN$.

For $\nu =2$, a direct evaluation of (\ref{fwalksagain})
leads to
\begin{equation}
{\cal N}^F(N,m,\nu) \sim 2^m,\;\; \nu=2,
\label{nu2}
\end{equation}
which is independent of $N$. This happens because 
at $\nu =2$ the walkers are ``densely packed'' and
can only move in unison, effectively behaving like
a single random walker.


\vspace{1cm}

\begin{thebibliography}{xx}

\bibitem{amb} J.\ Ambj\o rn:
{\it Simplicial Euclidean and Lorentzian quantum gravity},
plenary talk given at GR16 [gr-qc/0201028].

\bibitem{loll} R.\ Loll: {\it Discrete Lorentzian quantum gravity},
Nucl.\ Phys.\ B\ (Proc.\ Suppl.)\ 94 (2001) 96-107 [hep-th/0011194].

\bibitem{ajl1} J.\ Ambj\o rn, J.\ Jurkiewicz and R.\ Loll:
{\it A nonperturbative Lorentzian path integral for gravity},
{Phys.\ Rev.\ Lett.}\ 85 (2000) 924-927 [hep-th/0002050].

\bibitem{d3d4} J.\ Ambj\o rn, J.\ Jurkiewicz and R.\ Loll:
{\it Dynamically triangulating Lorentzian quantum gravity},
Nucl.\ Phys.\ B 610 (2001) 347-382 [hep-th/0105267].

\bibitem{ajl2} J.\ Ambj\o rn, J.\ Jurkiewicz and R.\ Loll:
{\it Nonperturbative 3d Lorentzian quantum gravity},
Phys.\ Rev.\ D 64 (2001) 044011 [hep-th/0011276].

\bibitem{ajl3} J.\ Ambj\o rn, J.\ Jurkiewicz and R.\ Loll:
{\it Computer simulations of 3d Lorentzian quantum gravity},
Nucl.\ Phys.\ B\ (Proc.\ Suppl.) 94 (2001) 689-692 [hep-lat/0011055].

\bibitem{adjl} J.\ Ambj\o rn, A.\ Dasgupta, J.\ Jurkiewicz and R.\ Loll:
{\it A Lorentzian cure for Euclidean troubles},
Nucl.\ Phys.\ Proc.\ Suppl.\ 106 (2002) 977-979 [hep-th/0201104].

\bibitem{ajl4} J.\ Ambj\o rn, J.\ Jurkiewicz and R.\ Loll:
{\it 3d Lorentzian, dynamically triangulated quantum gravity},
Nucl.\ Phys.\ Proc.\ Suppl.\ 106 (2002) 980-982 [hep-lat/0201013].

\bibitem{conf} A.\ Dasgupta and R.\ Loll: {\it A proper-time cure for the
conformal sickness in quantum gravity}, Nucl.\ Phys.\ B\ 606 (2001) 
357-379 [hep-th/0103186];
A.\ Dasgupta: {\it The real Wick rotations in quantum gravity},
preprint Golm AEI-2002-011, Feb 2002 [hep-th/0202018].

\bibitem{ajlv} J.\ Ambj\o rn, J.\ Jurkiewicz, R.\ Loll and G.\ Vernizzi:
{\it Lorentzian 3d gravity with wormholes via matrix models},
JHEP 0109 (2001) 022 [hep-th/0106082].

\bibitem{kz} V.A.\ Kazakov and P.\ Zinn-Justin:
{\it Two matrix model with ABAB interaction},
Nucl.\ Phys.\ B\ 546 (1999) 647-668 [hep-th/9808043].

\bibitem{al} J.\ Ambj\o rn and R.\ Loll:
{\it Non-perturbative Lorentzian quantum gravity, causality and
topology change},
Nucl.\ Phys.\ B\ 536 (1998) 407-434 [hep-th/9805108].

\bibitem{charlotte}P.\ Di Francesco, E.\ Guitter and C.\ Kristjansen:
{\it Integrable 2-d Lorentzian gra\-vi\-ty and random walks},
Nucl.\ Phys.\ B\ 567 (2000) 515-553 [hep-th/9907084];
{\it Generalized Lorentzian triangulations and the Calogero Hamiltonian},
Nucl.\ Phys.\ B\ 608 (2001) 485-526 [hep-th/0010259].

\bibitem{dehne} C.\ Dehne:
{\it Konstruktionsversuche eines quantenkosmologischen,
dynamisch triangulierten Torusuniversums in 2+1 Dimensionen}
(in German), Diploma Thesis, Univ. Hamburg (2001),
http://www.aei-potsdam.mpg.de/research/thesis/ dehne\_dipl.ps.gz.

\bibitem{2+1} see, for example, the textbook on
{\it Quantum gravity in 2+1 dimensions} by S.\ Carlip, Cambridge
University Press, Cambridge, UK (1998), and references therein

\bibitem{jormaetal} J.\ Louko and P.J.\ Ruback:
{\it Spatially flat quantum cosmology},
Class.\ Quant.\ Grav.\ 8 (1991) 91-121;
J.\ Louko and P.A.\ Tuckey:
{\it Regge calculus in anisotropic quantum cosmology},
Class.\ Quant.\ Grav.\ 9 (1992) 41-67;
D.\ Giulini and J.\ Louko:
{\it No boundary theta sectors in spatially flat quantum cosmology},
Phys.\ Rev.\ D\ 46 (1992) 4355-4364
[hep-th/9203007].

\bibitem{kappel} D.\ Kappel:
{\it Nichtperturbative Pfadintegrale der Quantengravitation durch
kausale dynamische Triangulierungen} (in German), Diploma Thesis,
Univ. Potsdam (2001).

\bibitem{dittrich} B.\ Dittrich: 
{\it Dynamische Triangulierung von Schwarzloch-Geometrien}
(in German), Diploma Thesis, Univ. Potsdam (2001).

\bibitem{fisher} M.E.\ Fisher:
{\it Walks, walls, melting, and wetting}, 
J.\ Stat.\ Phys.\ 34 (1984) 665-729.

\bibitem{essgutt}
J.W.\ Essam and A.J.\ Guttmann:
{\it Vicious walkers and directed polymer networks in general
dimensions}, 
Phys.\ Rev.\ E\ 52 (1995) 5849-5862.

\bibitem{guttvoe}
A.J.\ Guttmann and M.\ V\"oge:
{\it Lattice paths: vicious walkers and friendly walkers}, 
J.\ Statist.\ Plann.\ Inference\ 101 (2002) 107-131.

\bibitem{forr} 
P.J.\ Forrester:
{\it Exact solution of the lock step model of vicious walkers},
J.\ Phys.\ A\ 23 (1990) 1259-1273.

\bibitem{grab} D.J.\ Grabiner:
{\it Random walk in an alcove of an affine Weyl group, and 
non-colliding random walks on an interval},
J.\ Combin.\ Theory\ Ser.\ A\ 97 (2002) 285-306
[math.CO/0011218].

\bibitem{gz} I.M.\ Gessel and D.\ Zeilberger:
{\it Random walk in a Weyl chamber},
Proc.\ Amer.\ Math.\ Soc.\ 115 (1992) 27-31.

\bibitem{networks}
F.\ Markopoulou and L.\ Smolin,
{\it Causal evolution of spin networks},
Nucl.\ Phys.\ B\ 508 (1997) 409-430
[gr-qc/9702025].

\bibitem{twists} 
P.\ Peldan:
{\it Large diffeomorphisms in (2+1) quantum gravity on the torus},
Phys.\ Rev.\ D\ 53 (1996) 3147-3155
[gr-qc/9501020];
D.\ Giulini and J.\ Louko:
{\it Diffeomorphism invariant subspaces in Witten's 2+1 quantum
gravity on $R\times T^2$},
Class.\ Quant.\ Grav.\ 12 (1995) 2735-2746
[gr-qc/9504035];
S.\ Carlip and J.E.\ Nelson:
{\it The quantum modular group in (2+1)-dimensional gravity},
Phys.\ Rev.\ D\ 59 (1999) 024012
[gr-qc/9807087].

\bibitem{kuchryan} K.V.\ Kuchar and M.P.\ Ryan:
{\it Is minisuperspace quantization valid?: Taub in mixmaster},
Phys.\ Rev.\ D\ 40 (1989) 3982-3996.

\bibitem{bojo} M.\ Bojowald:
{\it Isotropic loop quantum cosmology},
preprint Penn State Univ. CGPG-02-2-2, Feb 2002
[gr-qc/0202077].


\end{thebibliography}
\end{document}
\end